\documentclass{elsarticle}

\usepackage{lineno,hyperref}

\usepackage{vdrcmac_withoutSubDir}
\usepackage{url}
\usepackage{mathtools}

\journal{Computer-Aided Design}

\usepackage{numcompress}

\begin{document}

\begin{frontmatter}

\title{Support-Free Hollowing for 3D Printing via Voronoi Diagram of Ellipses}

\author[address1]{Mokwon Lee\fnref{footnote1}}
\author[address2]{Qing Fang\fnref{footnote1}}
\fntext[footnote1]{Mokwon Lee and Qing Fang equally contributed.}
\author[address3]{Joonghyun Ryu}
\author[address2]{Ligang Liu\corref{corr}}
\ead{lgliu@ustc.edu.cn}
\author[address1,address3]{Deok-Soo Kim\corref{corr}}
\ead{dskim@hanyang.ac.kr}

\address[address1]{School of Mechanical Engineering, Hanyang University, Korea}
\address[address2]{School of Mathematical Science, University of Science and Technology of China, China}
\address[address3]{Molecular Geometry and Voronoi Diagram Research Center, Hanyang University, Korea}

\cortext[corr]{Corresponding author. Tel./Fax: +86-551-63600985 (Ligang Liu), Tel.: +82-2-2220-0472, Fax: +82-2-2292-0472 (Deok-Soo Kim)}

\begin{abstract}

Recent work has demonstrated that the interior material layout of a 3D model can be designed to make a fabricated replica satisfy
application-specific demands on its physical properties such as resistance to external loads. A widely used practice to fabricate such
models is by layer-based additive manufacturing (AM) or 3D printing, which however suffers from the problem of adding and removing interior
supporting structures.
In this paper, we present a novel method for generating support-free elliptic hollowing for 3D shapes which can entirely avoid additional supporting structures.
To achieve this, we perform the ellipse hollowing in the polygons on parallel section planes and protrude the ellipses of one plane to its neighboring planes.
To pack the ellipses in a polygon, we construct the Voronoi diagram of ellipses to efficiently reason the free-space around the ellipses and other geometric features by taking advantage of the available algorithm for the efficient and robust construction of the Voronoi diagram of circles.
We demonstrate the effectiveness and feasibility of our proposed method by generating interior designs for and printing various 3D shapes.

\end{abstract}


\end{frontmatter}


\section{Introduction}




Additive manufacturing (AM), also called 3D printing, has been advanced rapidly to make customized 3D models. Comparing to traditional manufacturing techniques, it offers enormous geometrical freedom for designers to create highly optimized components with various functionality.

Model hollowing is a typical practice for purposes of reducing printing material and time in 3D printing of light-weighted artifacts and various methods on generating optimized interior have been developed during the last few years~\cite{Stava:2012,Wang:2013,Lu:2014}.
The mainstream of 3D printing technologies, such as Fused Deposition Modeling (FDM) and Stereolithography (SLA), requires additional supporting structures to avoid the falling of relatively large overhanging parts during the printing process~\cite{Strano:2013,Dumas:2014,Vanek:2014}.
Generally the extra supporting structures have to be removed either manually or by dissolving dissolvable material from the printed objects.

However, there is no way at all to remove the supporting material inside interior voids of a printed object. A na\"{i}ve method is to first decompose the model into a few subparts, print them and remove the supporting material individually, and then glue them together. Obviously it may largely affect the computed physical properties.
Recently, there are quite a few attempts to create support-free interior voids or structures by constraining boundary slopes~\cite{Langelaar:2016,Wu:2016,Reiner:2016}.

A noticeable work adopts rhombic cell structures, where the slope angles of all rhombic cells are smaller than a prescribed maximum overhang-angle, as a infill pattern to generate support-free interior voids inside the objects~\cite{Wu:2016}.
However, the rhombic cells have only $C^0$ boundaries, which suffers serious problem of stress concentration~\cite{Walter:2008}. The stress around a discontinuity, e.g., a $C^0$ corner, will be excessively high when compared to the stresses at the other smooth areas as shown in Figure~\ref{fig:stress-concentration}.
For example, hatches and doorways in airplanes are oval to stay away from being broken easily~\cite{GeekStress:2016}.
Rounded corners are structurally more beneficial than sharp corners and also reduce the probability of crack development unlike sharp corners.

\FigOne{stress-concentration}{0.8}
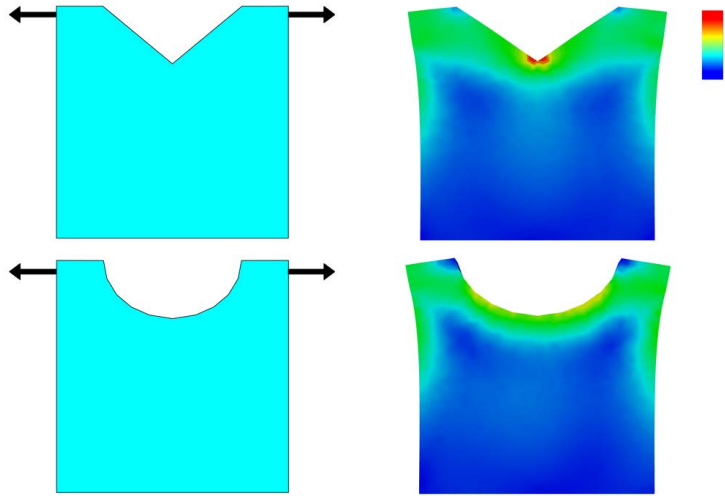
{Stress analysis on two thin plates by conducting two external loads (represented by two arrows respectively) on them. There is a $C^0$ discontinuity in the concave region of the upper shape while the concave region of the lower shape has a smooth boundary. From the stress maps, it is seen that the stress around the discontinuity is excessively high.}

\paragraph{Our Work}
To this end, we present a novel method for hollowing 3D shapes with support-free smooth elliptic interior voids.
To make it simple, we first derive a class of support-free ellipses in 2D case, based on the observation on sticky property of printing material.
Then we develop a novel approach for packing these support-free ellipses in the interior of 2D shapes, which is a very challenging problem.
To achieve this goal, we develop a greedy but efficient algorithm on adding these ellipses successively in 2D shapes via the Voronoi diagram of ellipses in polygons using that of circular disks.
Then the hollowed ellipses are extruded into 3D volume and thus a support-free hollowed 3D shape is generated.
Various experimental results have demonstrated the feasibility and applicability of our proposed method.

\paragraph{Contributions}
Our contributions are summarized as follows.
\begin{itemize}
    \item We develop efficient algorithms for computing the Voronoi diagram of polygon and that of ellipses within a polygon;
    \item We develop an efficient method for packing 2D ellipses with derivation of support-free constraint;
    \item We propose a method for generating support-free elliptic voids for 3D shapes.
\end{itemize}

To the best of our knowledge, our work is the first to offer a framework to generate a smooth, elliptic support-free interior voids for printing 3D shapes via an efficient computational scheme.
This provides a practically feasible operator for support-free hollowing of 3D shapes and exemplifies research along this direction.

\section{Related Work}

3D printing technology has drawn a lot of attention in geometric and physical modeling and optimization in computer graphics community.
We discuss about the work closely related to our study and give specific discussion about their strength and limitations.

\paragraph{Supporting Structures for 3D Printing}
Extra supporting structures are required to make 3D shapes printable for shapes with large overhanging parts, which leads to material waste and longer printing time.
Some methods, which adopt various supporting structures such as scaffold-like structures~\cite{Dumas:2014} and tree-like ones~\cite{Vanek:2014}, have been developed to generate economic usage of supporting structure for 3D models.
Vanek et al.~\cite{Vanek:2014} searches an optimal printing direction by reducing the total area of facing down regions where require additional supporting structures while Zhang et al.~\cite{Zhang:2015} develops a training-and-learning model to determine the optimal printing direction considering multiple factors such as contact area, viewpoint preference, and visual saliency.
The work of~\cite{Hu:2015} optimizes the shape of an input model to reduce the area of facing down regions for less supporting structures.
However, adding supporting structure for interior voids will suffer the serious problem that it is impossible to remove these extra structures without breaking the object into pieces.
Instead, we generate interior support-free voids, which completely avoids the usage of extra supports.

\paragraph{Interior Hollowing}
Significant work has been done in generating interior structure of a model to meet various geometric and/or physical properties.
Stava et al.~\cite{Stava:2012} hollows a 3D-printed object while maintaining its structural strength by adding some internal struts.
The interior is optimized by a reduced-order parameterization of offset surfaces in~\cite{Musialski:2015}.
Various internal structures, such as the skin-frame structure~\cite{Wang:2013}, the honeycomb-like structure~\cite{Lu:2014}, and the medial axis tree structure~\cite{Zhang:2015}, were developed for cost-effective purposes while preserving the structural strength of printed objects.
Both static balance and dynamic balance have been studied by designing the interior infills as well as changing the model shapes~\cite{Prevost:2013,Christiansen:2015,Bacher:2014}.
Instead of designing hollowing structure explicitly, a lot of efforts have been put on topology optimization to obtain distributions of material according to certain performance criteria during the last three decades~\cite{Deaton:2014,Wu:2016-TVCG}.
However, these works have not handled the problem of avoiding large overhangs. We study this problem by developing a carving operator via support-free elliptic voids.

\paragraph{Support-Free Structure}
Hu et al.~\cite{Hu:2014-Pyramidal} proposes a method to decompose a 3D object into support-free pyramidal subparts.
Reiner and Lefebvre~\cite{Reiner:2016} proposes an interactive sculpting system for designing support-free models.
Recent attempts have been put on creating support-free interior structure for 3D printing.
Wu et al.~\cite{Wu:2016} develops a method to generate support-free infill structures on adaptive rhombic cells.
A concurrent work of Langelaar~\cite{Langelaar:2016} considers an overhang angle threshold in topology optimization and generates a support-free material distribution.
However, these works only involve overhang angle and generate $C^0$ boundaries inevitably, resulting in large stress concentration in discontinuities.
In this paper, we develop a special class of support-free ellipses and adopt them as an interior carving operator to create infill structures.

\paragraph{Ellipse Packing}
Our method of ellipse carving is quite related to the problems of ellipse and ellipsoid packing which are NP-hard.
In science, the problem was prevalently approached from the packing ratio, particle size distribution, or jamming point of view to understand material properties by enforcing contacts between ellipses and using Monte Carlo method~\cite{BuchalterBradley92,SchreckEtal10}, Molecular Dynamics~\cite{DonevEtal04,DonevEtal07}, local and greedy algorithms~\cite{Hitti:2013,Birgin:2016} with sampling points on the ellipse boundary~\cite{DelaneyEtal05} in either regular containers or arbitrary domains~\cite{Lozano:2016}.
However, these methods do not balance the computation cost and packing density outcome well because research interests were primarily on the discovery of new phenomena.
This is quite different from the circle packing problem which have been extensively studied from the view points of both solution quality (i.e. packing ratio) and computation time, from early study using an event-driven algorithm with a bucket acceleration~\cite{Jodrey:1985,Lubachevsky:1990,Lubachevsky91} to recent ones using a systematic reasoning of empty space~\cite{Specht15,SugiharaEtal04}.
%
However, the infilled ellipses in our work have special constraints which make current methods infeasible to use.
In this work, we develop a new method for packing ellipses in arbitrary shapes with the mathematical tool of Voronoi diagram (VD).
Specifically, we perform an ellipse packing in the polygons on parallel section planes of a 3D model and protrude the ellipses of one plane to its neighboring planes in the shape volume to generate the hollowed results.
To better pack the ellipses in a polygon, we construct the Voronoi diagram of ellipses to efficiently reason the free-space around the ellipses and other geometric features by taking advantage of the available algorithm for the efficient and robust construction of the Voronoi diagram of circular disks which approximate the ellipses~\cite{LmwEtal16}.
The algorithm inherits the disk packing algorithm using the VD of disks~\cite{SugiharaEtal04}.

\section{Notations and Overview}

For the sake of simplicity, we first elaborate on our method in a 2D setting. The extension to 3D is realized by extrusion in Section~\ref{sec:extrusion}.

\subsection{Support-Free Ellipses}

\paragraph{Fabrication and Material Parameters}
Denote $\sigma_0$ as the printing precision, i.e., the thickness of each fabrication layer, which is $0.1-0.4 \texttt{mm}$ for general FDM printers.
Due to the sticky property of printing material, a short length $\delta_0$ of horizontal hangover can be successfully printed without extra supporting structure.
Denote $\theta_0$ as the maximally allowed overhang-angle.
The material-dependent parameters $\theta_0$ and $\delta_0$ can be measured by experiments, for example, $\theta_0=60^{\circ}$ and $\delta_0=5 \texttt{mm}$ for plastic PLA material.
The minimum wall thickness is set as $\delta = 5\delta_0$.

%
\paragraph{Support-Free Ellipses}
Denote $a$ and $b$ as the horizontal axis and vertical axis of an ellipse $E$, respectively.
A hollowed ellipse is called \emph{support-free} if it can be printed without any extra supporting structure.
Given an ellipse shown in Figure~\ref{fig:Ellipse-SupportFree}(a), denote $P_1$ and $P_2$ as the two points whose tangent lines have an overhang angle of $\theta_0$.
By many experiments we have proved the following observation: if the distance between $P_1$ and $P_2$ is no larger than $\delta_0$, then this ellipse is support-free, that is, the elliptic arc between $P_1$ and $P_2$ (shown in red) can be safely printed without any extra support.
On the other hand, too small interior ellipses cannot be printed and thus we set a lower bound of $a$ as $5\delta_0$.
Based on the observations, we have derived the conditions for a support-free ellipse as:
\begin{equation}\label{eq:EQSPFCondition}
  \begin{cases}
  b\ge a, \text{if}\ \ 5\sigma_0 \le a \le \frac{\delta_0}{2\cos\theta_0},\\
  b\ge a\frac{\sqrt{4a^2-\delta_0^2}}{\delta_0\tan\theta_0}, \text{if}\ \ a\ge \frac{\delta_0}{2\cos\theta_0}.
  \end{cases}
\end{equation}
%
%
It is worthwhile to mentioning that a uniform shrinking of a support-free ellipse is still support-free as the support-free conditions in Equation~\eqref{eq:EQSPFCondition} are convex.

\FigTwo{Ellipse-SupportFree}{0.35}{0.35}
{
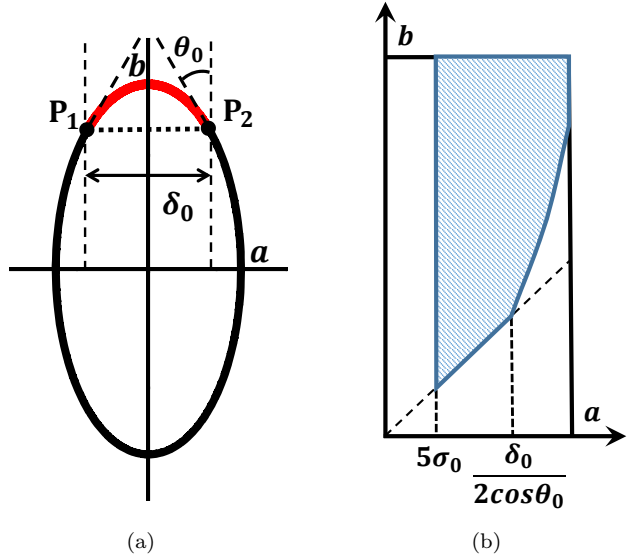
(a) The condition of a support-free ellipse: $\|P_1 P_2\| \le \delta_0$ where $P_1$ and $P_2$ are the two points with tangent lines of overhang angle $\theta_0$;
(b) Ellipses with $a$ and $b$ within the gray region are support-free.
}

\FigFour{VD-illustration}{0.47}{0.47}{0.47}{0.47}
{
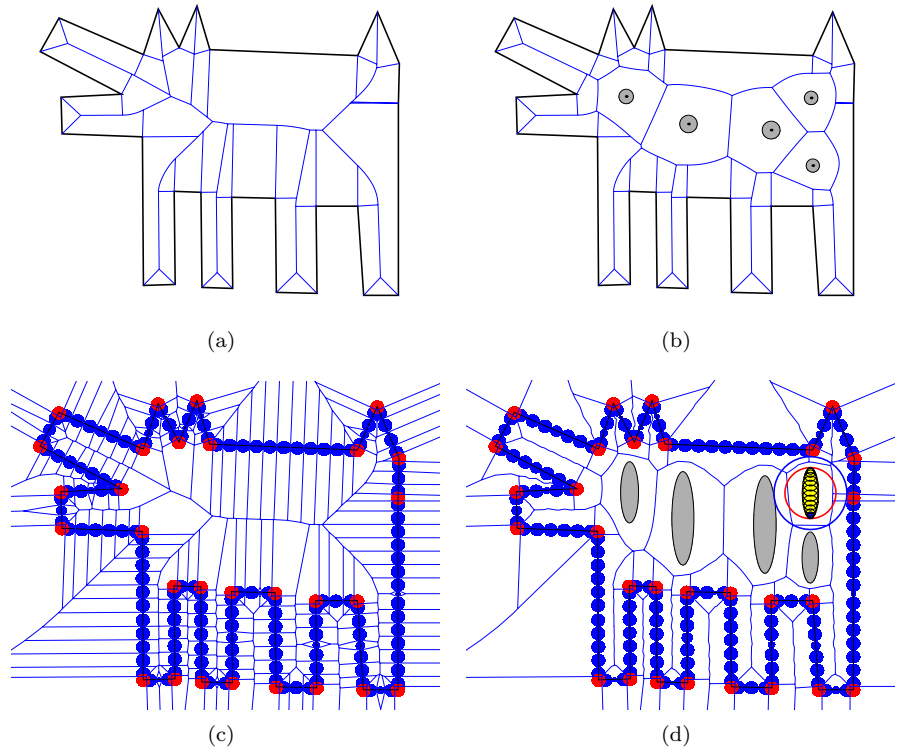
Illustration of VDs in this study.
(a) The VD of a polygon $\mathcal{P}$ ($\mathcal{VD}(\mathcal{P})$);
(b) The VD of a disk set $\mathcal{D}$ within $\mathcal{P}$ ($\mathcal{VD}(\mathcal{P}, \mathcal{D})$);
(c) The VD of the disk approximation $\tilde{\mathcal{P}}$ of $\mathcal{P}$ ($\mathcal{VD}(\tilde{\mathcal{P}})$);
(d) The VD of ellipses within $\tilde{\mathcal{P}}$ after the V-faces belonging to a P-edge are merged  ($\mathcal{VD}(\tilde{\mathcal{P}}, \mathcal{E})$).
}

\FigFive{idea-overview-5-bunnies}{0.38}{0.38}{0.38}{0.38}{0.38}
{
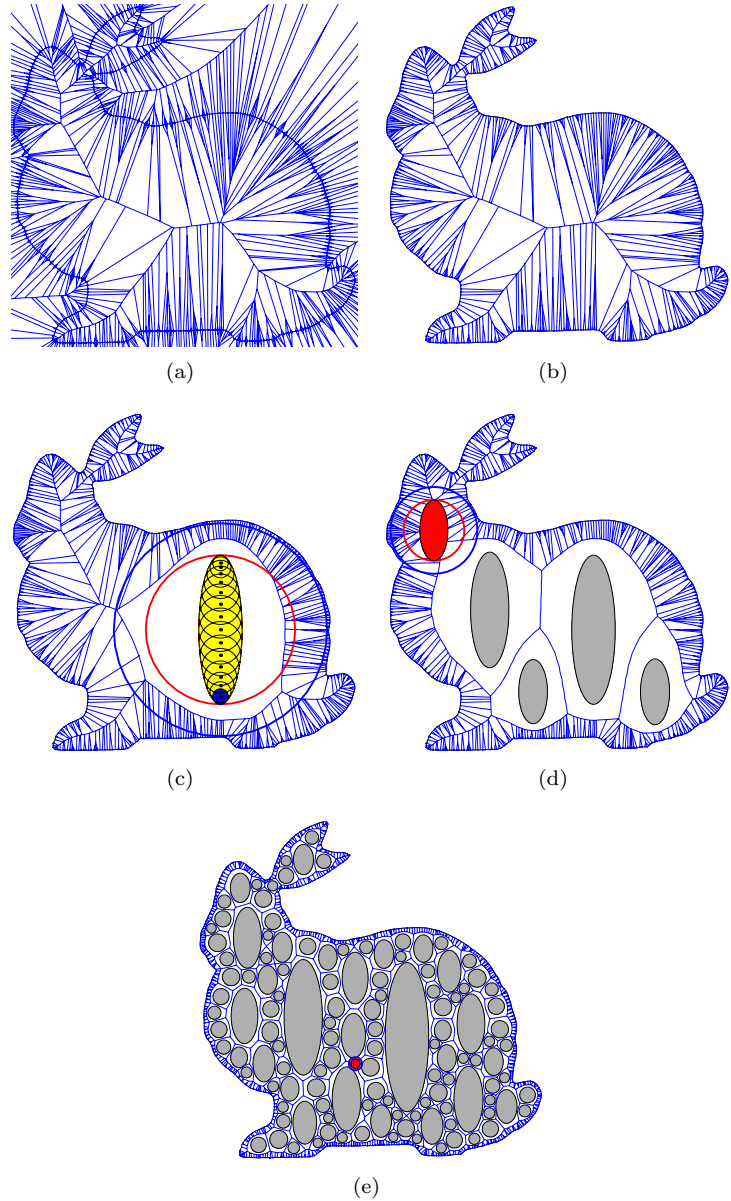
The ellipse packing process for the bunny polygon using the VD of ellipses.
(a) The VD of boundary disks.
(b) The VD of the bunny interior.
(c) The in-disks of the first ellipse within the clearance probe are incremented.
(d) The fifth ellipse located within the clearance probe after the four ellipses were incremented.
(e) The 100-th ellipse to be incremented.
}

\subsection{Ellipse Packing Method}

\paragraph{Problem}
Given a polygon $\mathcal{P}$, our goal is to find an optimal hollowing, i.e., packing, of a set of $m$ ellipses $\mathcal{E}=\{E_1,E_2,\ldots,E_m\}$ within $\mathcal{P}$ so as to maximize the sum of the areas all ellipses (the packing ratio) while satisfying mechanical and physicochemical constraints.

\paragraph{2D Polygon}
Given a 3D mesh $\mathcal{M}$ representing the boundary of an oriented solid, possibly with handles and interior voids, and a fabrication orientation ($z$-axis), we project $\mathcal{M}$ onto planes parallel to $z$-axis and choose the projection direction with the largest projected silhouette area.
Then $\mathcal{M}$ is represented as a sequence of parallel cross-sections, i.e., 2D polygonal shapes $\mathcal{P}^{*} = \{\mathcal{P}_1, \mathcal{P}_2, \cdots, \mathcal{P}_k \}$, along the chosen projection direction.
We choose the polygon in $\mathcal{P}^{*}$ with the largest area and denote it as $\mathcal{P}$ which may have internal holes (voids).
Denote $\mathcal{P}=(V,E)$ where $V$ and $E$ are the sets of vertices and edges, thus denoted as P-vertices and P-edges, respectively.
If $\mathcal{P}$ contains holes inside, its boundary is represented as a few closed loops: One outer loop (with counter-clockwise vertices) and a few inner ones (with clockwise vertices).
There can be more than one polygon on a section plane.
In the following, we present our method on hollowing $\mathcal{P}$ with support-free ellipses, i.e., packing of support-free ellipses in $\mathcal{P}$.

\paragraph{Voronoi Diagrams (VD)}
We use Voronoi diagrams (VD) to compute the hollowed ellipses because it is the most efficient and compact data structure for spatial reasoning among particles.
The VD of a generator set is the tessellation of space such that each cell of the tessellation consists of the locations closer to a corresponding generator than to others.
Various types of VD can be defined by generalizing the generator type, the distance definition, and the dimension.
For details, see \cite{OkabeEtal99}.

\paragraph{Voronoi Diagrams of Disks and Ellipses}
In this study, we use the VDs of disks and ellipses within a polygon $\mathcal{P}$ with the Euclidean $l_2$-distance in 2D.
See Figure~\ref{fig:VD-illustration}:
(a) The VD of the polygon $\mathcal{P}$ ($\mathcal{VD}(\mathcal{P})$);
(b) The VD of a disk set $\mathcal{D}$ within $\mathcal{P}$ ($\mathcal{VD}(\mathcal{P}, \mathcal{D})$);
(c) The VD of the disk approximation $\tilde{\mathcal{P}}$ of $\mathcal{P}$ ($\mathcal{VD}(\tilde{\mathcal{P}})$);
(d) The VD of ellipses within the approximation $\tilde{\mathcal{P}}$ ($\mathcal{VD}(\tilde{\mathcal{P}}, \mathcal{E})$).
Note that both $\mathcal{VD}(\mathcal{P})$ and $\mathcal{VD}(\mathcal{P}, \mathcal{D})$ can be correctly, efficiently, and robustly computed.
However, we compute $\mathcal{VD}(\tilde{\mathcal{P}}, \mathcal{E})$ instead of $\mathcal{VD}(\mathcal{P}, \mathcal{E})$ because of the challenges involved in the computation of V-vertices and V-edges of $\mathcal{VD}(\mathcal{P}, \mathcal{E})$, which will be explained in detail in Section~\ref{sec:extrusion}.
Hence, we construct the approximation $\mathcal{VD}(\tilde{\mathcal{P}})$ instead of $\mathcal{VD}(\mathcal{P})$ where each P-vertex is associated with a disk called at-disk (the red filled-circles in Figure~\ref{fig:VD-illustration}(c,d)) and each P-edge is associated with more than two disks called on-disks (the blue filled-circles).
In Figure~\ref{fig:VD-illustration}(d), the blue thick circle is the maximum clearance probe $\pi_{max}$, centered at a V-vertex of $\mathcal{VD}(\tilde{\mathcal{P}}, \mathcal{E})$, that guarantees its inscribing ellipse is intersection-free with any other existing ellipses and the red thick circle is 70\% shrunken probe $\tilde{\pi}_{max}$ in which the actual ellipse is created.
The shrinking ratio is given by users according to their intention to control the overall distribution of ellipse heights.
Be aware that $\tilde{\mathcal{P}}$ is associated with a disk set $\mathcal{D}^{\mathcal{P}}=\{\mathcal{D}_{at},\mathcal{D}_{on}\}$ where $\mathcal{D}_{at}$ and $\mathcal{D}_{on}$ are the sets of at-disks and on-disks, respectively.

\paragraph{Idea of the Packing Algorithm}
Let $\mathcal{VD}(\tilde{\mathcal{P}})$ be the VD of the interior of $\mathcal{P}$ (Figure~\ref{fig:idea-overview-5-bunnies} (b)) obtained by trimming the exterior part of the entire VD in Figure~\ref{fig:idea-overview-5-bunnies} (a) and $\mathcal{VD}(\tilde{\mathcal{P}},\mathcal{E})$ that of $\mathcal{E}$ within $\mathcal{P}$.
Let $\pi_v$ be the biggest empty circle centered at a V-vertex $v$ of $\mathcal{VD}(\tilde{\mathcal{P}})$ where its radius is given by the distance from $v$ to the boundary of its generators.
Let $\pi_{max}$ be the possible biggest one, called the maximum clearance probe, among all such empty circles from the V-vertices of $\mathcal{VD}(\tilde{\mathcal{P}})$.
Let $v_{max}$ be the V-vertex corresponding to $\pi_{max}$.
%
Starting from $\mathcal{E}=\{\emptyset\}$, we first find the V-vertex $v_{max}$ with $\pi_{max}$ (the blue solid circle in Figure~\ref{fig:idea-overview-5-bunnies}(c)) and its shrunken probe $\tilde{\pi}_{max}$ (the red one) and place one new ellipse $E$ within $\tilde{\pi}_{max}$ (Figure~\ref{fig:VD-illustration}(d) and~\ref{fig:idea-overview-5-bunnies}(c)).
Then, we construct $\mathcal{VD}(\tilde{\mathcal{P}},\mathcal{E} \cup \{E\})$ by inserting $E$ into $\mathcal{VD}(\tilde{\mathcal{P}},\mathcal{E})$.
We repeat the clearance-probe-finding, the ellipse-placement, and Voronoi diagram update processes for a sufficient number of times until a termination condition is met, as shown in Figure~\ref{fig:idea-overview-5-bunnies}(d,e).

\section{Ellipse Hollowing via Voronoi Diagram}

\paragraph{Problem}
There are two major computational phases for using VD in the ellipse hollowing:
first, the construction of $\mathcal{VD}(\mathcal{P})$ and
second, the construction of $\mathcal{VD}(\mathcal{P},\mathcal{E})$.

\subsection{Voronoi Diagram of a Polygon}

\paragraph{Challenges}
It is well-known that the algorithm for an efficient and robust construction of $\mathcal{VD}(\mathcal{P})$ is \emph{not} trivial due to the influence of numerical error on maintaining correct topology of Voronoi diagram \cite{KdsEtal95,OkabeEtal99}.
In this study we developed and implemented a new, simple, and efficient yet robust algorithm based on the topology-oriented approach \cite{SugiharaIri92,Held01,HeldHuber09,LmwEtal16} for $\mathcal{VD}(\mathcal{P})$.
This is because we eventually need to construct $\mathcal{VD}(\mathcal{P},\mathcal{E})$ which is lacking in existing codes such as CGAL \cite{CGALHome,AlliezEtal07} and VRONI \cite{Held01}.

\paragraph{TOI-D Algorithm}
The proposed algorithm takes advantage of the Voronoi diagram of circular disks \cite{KdsEtal01,KdsEtal01b}, particularly the recently reported topology-oriented incremental (TOI) algorithm for computing the Voronoi of circular disks, thus abbreviated as the TOI-D algorithm, which takes $O(n^2)$ time in the worst case but $O(n)$ time on average for $n$ disks \cite{LmwEtal16}.
The idea is to approximate target geometric entities using circular disks in a sufficient resolution, construct the VD of the disks using the TOI-D algorithm, and merge some V-cells.
While a similar idea was used to curved objects using the ordinary Voronoi diagram of points which were sampled from curves \cite{Sugihara93,Held01,EmirisEtal13}, the proposed algorithm using the TOI-D algorithm is much powerful as circles can significantly reduce problem size and complexity.

\paragraph{Approximating Polygon with Disks}
We represent a polygon $\mathcal{P}=(V,E)$ as follows.
Let $e^* \in E$ be the shortest P-edge with its length $L^*$.
Suppose that we cover $e^*$ with two open disks with the diameter $L^*/2$ by placing their centers on $e^*$ and the boundary of each disk coincides either one of the two extreme points of $e^*$.
For each of the other P-edges with the length $L \ge L^*$, we place $\lfloor 2L/L^* \rfloor$ non-overlapping open disks in a sequel on the P-edge
($\lfloor · \rfloor$ denotes a floor function).
The uncovered remaining segment with the length $L - \lfloor 2L/L^* \rfloor L^*/2$ on the P-edge is then covered by one, and only one, smaller open disk.
We place this smaller disk in the middle of the P-edge for algorithmic simplicity.
Hence, the P-edge is entirely covered by $\lfloor 2L/L^* \rfloor + 1$ mutually exclusive open disks called an on-disk (The blue ones in Figure~\ref{fig:VD-illustration}(c)).
A disk $d$ is a child of an associated P-edge $e$ and $e$ the parent of $d$.
We also place a same size disk, called at-disk, $d_v$ at each P-vertex $v$ (The red ones in Figure~\ref{fig:VD-illustration}(c)): $v$ and $d_v$ also have a child-parent relationship.
Each disk knows its parent and each parent knows its children disks via pointers.
We eventually have a set $\mathcal{D}^{\mathcal{P}}$ of children disks representing $\mathcal{P}$ where no disk contains any other while two disks may intersect, $|\mathcal{D}^{\mathcal{P}}|>n$.

The computation speed of the idea above, which is sensitive to the shortest P-edge, can be improved by enforcing fewer on-disks for each P-edge.
We initially allocate only three on-disk: two near the extreme points of the P-edge with the radii identical to the at-disks for P-vertices and the third with the diameter covering the entire rest segment of the P-edge.
If the third on-disk intersects any other disk, either an at-disk or on-disk, we subdivide it (to avoid computational complications) until the intersection is resolved by employing a bucket system to accelerate the intersection check.
In this way, the number of children disks can be significantly reduced.
This approach of using on-disks with non-uniform sizes works well for the polygons produced from fine mesh models such as bunny.

However, there are models such that $L^* \gg \delta$ for the the minimum wall thickness $\delta$.
In such a case, the proposed algorithm both leaves an undesirable bulk material in printed artifacts and may run into a computational complication and thus we subdivide each P-edge into multiple shorter P-edges with $2\delta$.
Some engineering models with $L^* < \delta$ may have frequently large planar facets possibly with smaller ones.
We also subdivide long P-edges into several short ones with the same rule above.

\FigSix{VD-illustration-further}{0.44}{0.216}{0.44}{0.216}{0.44}{0.216}
{
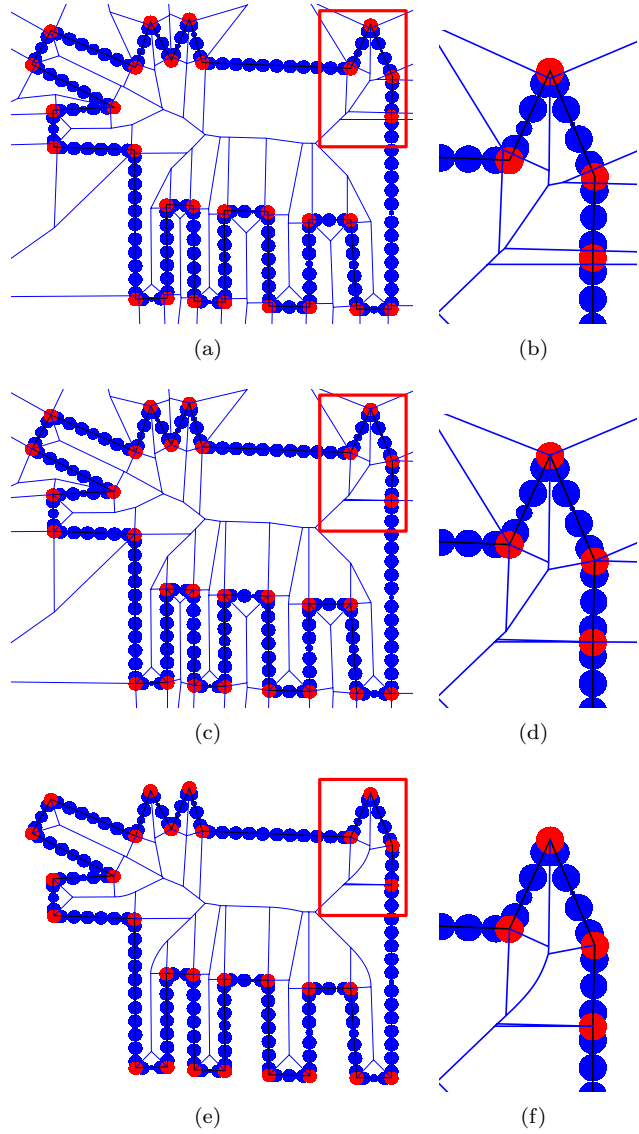
Three important steps of the TOI-P algorithm for $\mathcal{VD}(\mathcal{P})$.
(a,b) $\mathcal{VD}(\mathcal{D}^{\mathcal{P}})$ after the merge process.
(c,d) After some V-vertices are relocated.
(e,f) After outside V-edges are trimmed, the V-vertex coordinates and V-edge equations are computed.
}


\paragraph{TOI-P Algorithm}
We construct $\mathcal{VD}(\mathcal{D}^{\mathcal{P}})$ of the disk set $\mathcal{D}^{\mathcal{P}}=\{\mathcal{D}_{at},\mathcal{D}_{on}\}$ where $\mathcal{D}_{at}$ and $\mathcal{D}_{on}$ using the TOI-D algorithm (Figure \ref{fig:VD-illustration}(c)) and merge the V-cells of the children on-disks of each P-edge (Figure \ref{fig:VD-illustration-further}(a,b)).
As shown from the figure, the resulting VD structure has a unique V-cell for both each P-edge and each P-vertex and thus its structure is close to $\mathcal{VD}(\mathcal{P})$ from both topology and geometry point of views.
However, some V-vertices are off P-vertices which play the role of V-vertices (Figure \ref{fig:VD-illustration-further}(b)).
Hence, they should be moved to the related P-vertices.
Figure \ref{fig:VD-illustration-further}(c,d) shows the VD after relocating those V-vertices to the polygon boundary and flipping some V-edges to get the correct topology, both taking at most $O(m)$ time for $m$ disks.
Note that a V-edge flipping is required to get the correct topological structure of $\mathcal{VD}(\mathcal{P})$ (Compare Figure \ref{fig:VD-illustration-further}(b) and (d)).
Then, removing the exterior part of the VD and computing the V-vertex coordinates and the V-edge equations transforms the intermediate VD structure to the correct $\mathcal{VD}(\mathcal{P})$ (Figure \ref{fig:VD-illustration-further}(e,f)).
Note that some previously linear V-edges are curved.
Removing the at-disks and on-disks results in the VD of Figure \ref{fig:VD-illustration}(a).

There are three cases of V-edges:
i) If a V-edge $e$ is defined between two P-vertices, $e$ is a line segment;
ii) Between two P-edges, $e$ is also a line segment;
iii) Between a P-vertex and P-edge, $e$ is a parabolic arc.
As the V-edges of $\mathcal{VD}(\mathcal{P})$ are quadratic curve segments, they can be represented as a rational quadratic B\'{e}zier curve \cite{KdsEtal95}.
There are four cases of V-vertices:
i) Among three line segments;
ii) Among two line segments and one point;
iii) Among one line segment and two points;
iv) Among three points.
Each V-vertex coordinate and V-edge equation can be correctly computed in $O(1)$ time \cite{KdsEtal95}.

\paragraph{Time Complexity}
The TOI-D algorithm for the construction of the VD of $m$ disks takes $O(m)$ time on average and $O(m^2)$ time in the worst case \cite{LmwEtal16}.
With the polygon $\mathcal{P}$ of $n$ P-vertices and $n$ P-edges, $n<m$, the merge process takes $O(m-n)$ time because each merge of two adjacent V-cells takes $O(1)$ time.
As V-vertex shift and coordinate correction takes $O(n)$ time, the entire process for $\mathcal{VD}(\mathcal{P})$ takes $O(m)$ time provided that $\mathcal{VD}(\mathcal{D})$ is given.
As $m$ depends either on the length of the shortest P-edge or on the minimum wall thickness, the proposed TOI-P algorithm is input-sensitive.
All time complexities in this paper are in the worst case sense unless otherwise stated.

\subsection{Voronoi Diagram of Ellipses in a Polygon}

Regarding $\mathcal{VD}(\tilde{\mathcal{P}})$ for $\mathcal{D}^{\mathcal{P}}$ as $\mathcal{VD}(\tilde{\mathcal{P}},\mathcal{E}=\{\emptyset\})$, we first find the V-vertex $v_{max}$ with the maximum clearance probe $\pi_{max}$ and place an ellipse $E$ which inscribes $\tilde{\pi}_{max}$ so that its center coincides $v_{max}$.
We incrementally insert $E$ at $v_{max}$ of $\mathcal{VD}(\tilde{\mathcal{P}},\mathcal{E})$ to get $\mathcal{VD}(\tilde{\mathcal{P}},\mathcal{E} \cup \{E\})$.
Instead of directly inserting $E$ into $\mathcal{VD}(\tilde{\mathcal{P}},\mathcal{E})$, we insert each in-disk, one by one, into $\mathcal{VD}(\mathcal{D})$ where $\mathcal{D} \equiv \mathcal{D}^{\mathcal{P}}$.
As the ellipse increment process goes on, $\mathcal{D}$ contains all the in-disks of the ellipses incremented so far in addition to $\mathcal{D}^{\mathcal{P}}$.

\paragraph{Challenges}
Construction of $\mathcal{VD}(\mathcal{P},\mathcal{E})$ involves the computation of V-vertices and V-edges and the topological structure among them where each of these tasks is challenging.
Consider a V-vertex defined by three ellipses.
It is known that there can be up to 184 complex circles that are simultaneously tangent to three conics in the plane and each corresponds to a root of a polynomial of degree 184 \cite{EmirisTzoumas05}.
It is hard to expect to find the roots of a polynomial of such a high degree both exactly and efficiently.
Given 10-bit precision to represent the coefficients of three random ellipses, each coefficient of the resultant necessary for the exact computation of a V-vertex $v$ is, on average, 4603-bit integers \cite{EmirisTzoumas05}.
Hence, the exact and efficient computation of the correct coordinate of $v$ itself is not an easy problem at all.
We are not aware of any method to solve this resultant exactly and efficiently and thus an exact computation approach to construct the VD of ellipses seems impractical.
The V-edge between two ellipses can be more complicated than one might expect.
Even the bisector between a point and an ellipse can be very complicated \cite{FaroukiJohnstone94}:
It may have cusps and self-intersections and is disconnected if the point is located outside the ellipse.
The bisector between two rational curves can be non-rational and even a two-dimensional object \cite{AltEtal05}.
Therefore, the computation of V-vertices, V-edges, and the their association through the topological structure among ellipses in a free-space is a challenge, not to mention about the ellipses within a polygon.
As far as we know, no study has been reported for constructing $\mathcal{VD}(\mathcal{P},\mathcal{E})$.

\paragraph{Approximating an Ellipse with Circles}

%
We represent an ellipse $E$ as an approximation with a set $\mathcal{D}^E$ of an odd number of disks, called in-disks which inscribes $E$ (The yellow ones in Figure~\ref{fig:VD-illustration}(d) and Figure~\ref{fig:idea-overview-5-bunnies}(c)).
In-disks may intersect but none is contained by another.
The in-disks are generated as follows.
The first in-disk $d_1$ is the maximal inscribing disk which is centered at the center of $E$.
Let $\epsilon$ be an approximation error defined as the horizontal distance between $\partial E$ and $\partial d_1$.
Then, a point $p \in \partial d_1$ can be located for an \emph{a priori} defined error, say $\epsilon_0$.
Hence, a second in-disk $d_2$ passes through $p$ while inscribing $E$.
We alternate this calculation up and down of $d_1$ to get $d_2$ and $d_3$, respectively.
Repeating this calculation produces in-disks with a strictly controlled error bound $\epsilon_0$.
Given $\epsilon_0$, a shorter ellipse has fewer in-disks than a longer one does.


\paragraph{TOI-EinP Algorithm}

The idea is very simple as follows.
We insert each in-disk in $\mathcal{D}^E$ into $\mathcal{VD}(\mathcal{D})$ of existing disks and then merge the V-cells of the in-disks of $\mathcal{D}^E$.
If a sufficient number of in-disks approximates each ellipse, the VD of the disks well-approximate the VD of ellipses from both topology and geometry point of views.
As ellipses do not intersect, the in-disks from distinct ellipses do not intersect and the in-disks do not intersect both on-disks and at-disks on the polygon boundary, either.

The increment of an in-disk is done using the TOI-D algorithm.
When we increment an in-disk, we maintain a dual representation of both $\mathcal{VD}(\tilde{\mathcal{P}},\mathcal{E})$ and $\mathcal{VD}(\mathcal{D})$.
In other words, $\mathcal{VD}(\tilde{\mathcal{P}},\mathcal{E})$ and $\mathcal{VD}(\mathcal{D})$ are carefully synchronized in the following sense.
We first incrementally update $\mathcal{VD}(\mathcal{D})$ until all in-disks of in $\mathcal{D}^E$ (thus those of $E$) are exhausted to get $\mathcal{VD}(\mathcal{D} \cup \mathcal{D}^E)$.
Then, we merge the V-cells of the in-disks of $E$ to produce the topology of $\mathcal{VD}(\tilde{\mathcal{P}},\mathcal{E} \cup \{E\})$.
After the merge process, each of the remaining V-vertices of $\mathcal{VD}(\mathcal{D} \cup \mathcal{D}^E)$ becomes the V-vertices of $\mathcal{VD}(\tilde{\mathcal{P}},\mathcal{E})$ and a connected subset of some appropriate remaining V-edges becomes the V-edge of $\mathcal{D}^{\mathcal{P}}$.
Hence, we carefully maintain the correspondence of the V-vertices and V-edges between $\mathcal{VD}(\mathcal{D} \cup \mathcal{D}^E)$ and $\mathcal{VD}(\tilde{\mathcal{P}},\mathcal{E} \cup \{E\})$.
Note that the merge can also be done incrementally as soon as after an in-disk is incremented.
Figure \ref{fig:sequential-illustration-hollow-6-dogs} shows the process of incrementing ellipses:
(a) The clearance probe $\pi_{max}$ (the large blue circle), its shrunken probe $\tilde{\pi}_{max}$ (the red circle), the ellipse within $\tilde{\pi}_{max}$, and the biggest in-disk (blue filled circle) is incremented into the VD;
(b) The second in-disk (the blue filled circle) is incremented into the VD (The previously incremented in-disk is yellow now);
(c) After four in-disks are incremented;
(d) After all in-disks of the first ellipse are incremented;
(e) After the second ellipse is incremented;
(f) After the third ellipse is incremented.

The V-vertices of $\mathcal{VD}(\tilde{\mathcal{P}},\mathcal{E} \cup \{E\})$ remaining after the V-cell merge have their coordinates inheriting from $\mathcal{VD}(\mathcal{D} \cup \mathcal{D}^E)$ which are computed from a triplet of in-disks.
Thus, they are not necessarily correct for ellipses and it is necessary to compute their correct coordinates for the successful packing of next ellipse because the maximum clearance probe needs to be found from these V-vertices.
The six cases of generator combination for a V-vertex in $\mathcal{VD}(\mathcal{P}, \mathcal{E} \cup \{E\})$ among ellipses, line segments (i.e. P-edges), and points (i.e. P-vertices) becomes a unified case of among three ellipses in $\mathcal{VD}(\tilde{\mathcal{P}}, \mathcal{E} \cup \{E\})$ because of $\mathcal{D}_{at}$ and $\mathcal{D}_{on}$.
The six cases are as follows:
among three ellipses,
among two ellipses and one line segment,
among two ellipses and one point,
among one ellipse and two line segments,
among one ellipse and two points, and
among one ellipse, one line, and one point.

\FigSix{sequential-illustration-hollow-6-dogs}{0.384}{0.384}{0.384}{0.384}{0.384}{0.384}
{
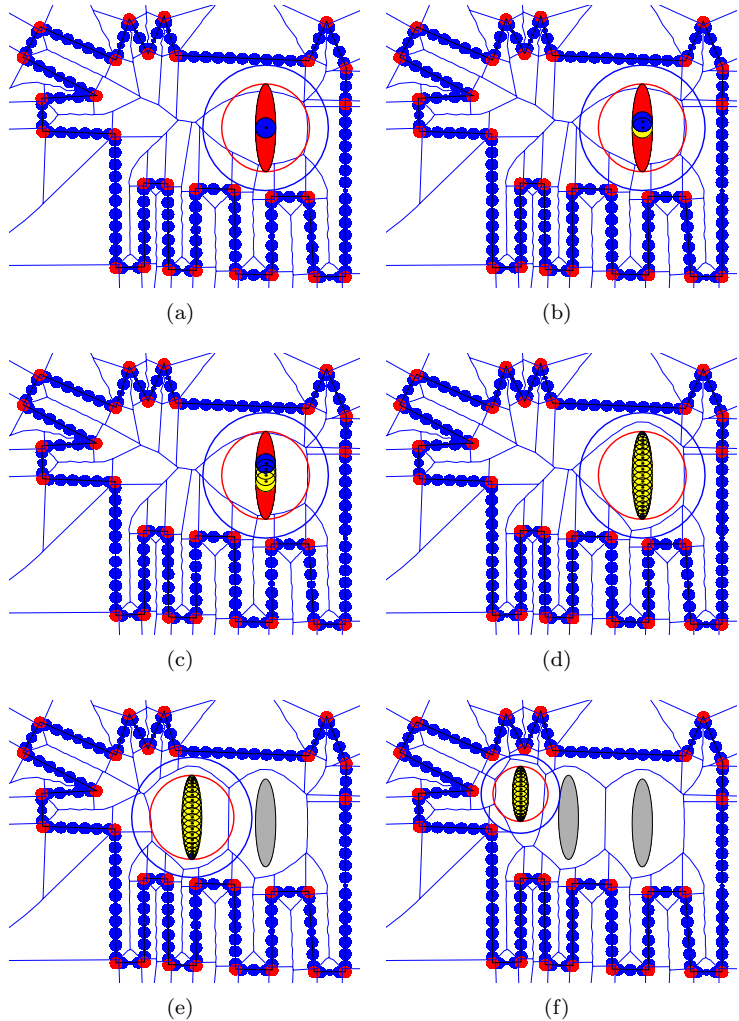
The ellipse increment process of the TOI-EinP algorithm through the increments of in-disks into the VD structure.
(a) Identification of the clearance probes, the ellipse, and the increment of the biggest in-disk.
(b) the increment of the second in-disk.
(c) the increment of the fourth in-disk.
(d) after the increment of all in-disks of the first ellipse.
(e) after the increment of the second ellipse.
(f) the increment of the third ellipse.
}

\paragraph{V-vertex Coordinate among Three Ellipses}
Consider a V-vertex $v$ defined by three ellipse generators.
We iteratively find the correct location of $v$ in $\mathcal{VD}(\tilde{\mathcal{P}},\mathcal{E})$ starting with its initial coordinate provided by $\mathcal{VD}(\mathcal{D})$.
In other words, $v$ is initially equidistant from three in-disks where each is a child of each of three ellipses.
We project $v$ to each of the three ellipses to find its footprint (which is the closest location on an ellipse from $v$).
Then, we compute the circumcircle, say $\xi$, which passes through the three footprints and use the center of $\xi$ as the new coordinate of $v$.
Provided that each ellipse is approximated by a sufficient number of in-disks, the iteration of this footprint-projection and circumcircle-finding process quickly converges to the correct coordinate of $v$ due to the convexity of ellipse.
Experiment shows that the initial coordinate of $v$ is already very close to the converged coordinate.
The situation that a generator(s) of $v$ is at-disk or on-disk can be handled as an easier special case

\paragraph{V-edge Between Two Ellipses}

Due to the current theoretical limitations, it is practically inevitable to approximate a V-edge with a sequence of passing points computed through the envelopes of families of point/curve bisectors \cite{FaroukiJohnstone94,FaroukiJohnstone94b} or a sequence of curve segments \cite{FaroukiRamamurthy98}.
There exist other approaches to trace bisectors for VD and medial axis transformations \cite{OmirouDemosthenous06,CaoLiu08,CaoEtal09}.

We emphasize that the topological structure of $\mathcal{VD}(\mathcal{P},\mathcal{E})$ is already known.
In other words, for each V-edge $e$, its starting and ending V-vertices are known with correct coordinates along with its two elliptic generators.
We approximate each V-edge as a sequence of points by tracing the V-edge in a way conceptually similar to the tracing algorithm of the intersection curve between two free-form surfaces \cite{BarnhillEtal87}.
Tracing V-edges in this study is, however, much simpler than tracing general intersection curve in that i) the coordinates of two V-vertices of each V-edge are known, ii) V-edges are planar, and iii) each V-edge is $C^1$-continuous between two V-vertices.
The case that $e$ is defined between one ellipse and one at-disk (or on-disk) is an easier special case.

\paragraph{Finding Footprints}

Finding footprints is a key building block.
Suppose that $p$ is a point outside an ellipse $E$ and $L$ is a line passing through $p$.
It is known that there are four, three, or two locations on $E$ that $L$ perpendicularly intersects $E$ depending on whether $p$ lies inside the evolute, lies on the evolute but not at a cusp, or lies on a cusp or outside the evolute, respectively \cite{EmirisTzoumas05}.
Finding the perpendicular intersection between $L$ and $E$ can be formulated as a root-finding problem of a quartic polynomial thus taking $O(1)$ time.
The footprint of $p$ is obviously one of these locations which determines the minimum distance.

\paragraph{Time Complexity}
Suppose that $\mathcal{VD}(\tilde{\mathcal{P}},\mathcal{E})$ has $n$ elements of $P$ and $m_{\mathcal{E}}$ ellipses and suppose that $\mathcal{D}^{\mathcal{P}}$ has $N$ children disks and there are $M$ in-disks for all incremented ellipses so far.
Given a new ellipse $E$, with $C$ in-disks in $\mathcal{D}^E$, the increment of all in-disks into $\mathcal{VD}(\mathcal{D})$ takes $O(C(N+N))$ time.
Then, it is followed by the merges of the V-cells among the in-disks of $E$ taking $O(C)$ time.
This completes the computation of the topological structure of $\mathcal{VD}(\tilde{\mathcal{P}},\mathcal{E} \cup \{E\})$.
Then, the computation of the geometry of each of $O(n + m_{\mathcal{E}})$ V-vertices and V-edges takes $O(1)$ time.
Hence, given the synchronized $\mathcal{VD}(\tilde{\mathcal{P}},\mathcal{E})$ and $\mathcal{VD}(\mathcal{D})$, the increment of an ellipse takes $O(C(N+M) + n + m_{\mathcal{E}})$ time.
As $N \gg n$ and $M \gg m_{\mathcal{E}}$, the increment of one ellipse takes $O(N+M)$ time in the worst case.
Hence, the increment of $m_{\mathcal{E}}$ ellipses takes $O(m_{\mathcal{E}}N + C m_{\mathcal{E}}^2))$ time.

\subsection{Implementation Issues}

\paragraph{Criteria of Adding Ellipses}
During the ellipse increment, the distance between two adjacent interior ellipses should be no less than the minimum wall thickness $\delta = 5 \delta_0$.
%
An intuitive scheme is to maximally pack the polygon with ellipses and shrink each ellipse by $\delta/2$.
As an ellipse is not offset-invariant, the shrunk ellipse needs to be approximated but can be effectively computed.

We instead use a computationally easier yet equally effective scheme as follows.
We have two parameters to control the height of each ellipse: the minimum wall thickness $\delta$ and the shrink ratio $\rho \in (0,1)$ of the maximal clearance probe $\pi_{max}$.
Given $\pi_{max}$, we multiply $\rho$ to $\pi_{max}$ to get a shrunken probe $\tilde{\pi}_{max}$ with the shrunken radius $\gamma$.
If $\gamma > \delta/2$, we use $\tilde{\pi}_{max}$ to produce an inscribing ellipse and increment in the VD.
Otherwise, we reduce the radius of $\pi_{max}$ by $\delta/2$ and produce the inscribing ellipse.
The purpose of $\rho$ is to provide users a convenient handle to control the overall distribution of ellipse heights because, in addition to the NP-hardness of the optimal ellipse packing, we never know which way is best even if we only consider geometry.
Moreover, experienced users may want to have a control of overall shape distribution by tuning $\rho$.

\paragraph{Terminating Condition}

An ellipse should not be too small to be printed. We regard $a < \delta = 5 \delta_0$ (see Equation~\ref{eq:EQSPFCondition}) as the terminating condition of inserting new ellipses.
In other words, whenever $\pi_{max}$ is produced and shrunken, we check its horizontal axis $a$ and terminate if it is less than $\delta$.

\section{Extrusion to 3D}
\label{sec:extrusion}


\subsection{3D Hollowing}

\paragraph{Extrusion of Ellipses}
The hollowed polygon $\mathcal{P}^{*}$  is lifted to 3D by extruding the ellipses orthogonal to the 2D plane in both directions.
Denote $i$ as the index of $\mathcal{P}^{*}$, i.e., $\mathcal{P}_{i} = \mathcal{P}^{*}$.
For each ellipse $E$ in $\mathcal{P}_{i}$, we project it onto $\mathcal{P}_{i+1}$. If $E$ totally lies in $\mathcal{P}_{i+1}$ within a distance of minimal wall thickness $\sigma$, we keep it in $\mathcal{P}_{i+1}$. Otherwise, we shrink it with a factor so as to the shrunk ellipse lies in $\mathcal{P}_{i+1}$ within a distance of $\sigma$.
We can also enlarge the ellipse if there is much space around it. Note that the enlargement of the ellipse should meet the support-free condition (Equation~\ref{eq:EQSPFCondition}).
This operation is successively applied for the other cross sectional polygons.


\paragraph{Hollowing Other Polygons}

After we complete the extrusion of all ellipses for all cross sectional polygons, we check each polygon and choose the one with the largest available region which can insert more ellipses. Then we set it as input and add more support-free ellipses in it using our method, and then extrude the newly-added ellipses to its neighborhood polygons.
The above process is iteratively performed until there is no any ellipse which can be added into the polygons.

\paragraph{Hollowed Volume}
After we obtain the hollowed polygons, we connect the corresponding ellipses on successive polygons and thus generate a hollowed volume of $\mathcal{M}$. As each polygon is support-free, the obtained hollowed 3D volume is also support-free.
%


\subsection{Functional Constraints}

The printed objects are generally required to meet some functional constraints such as static balance and mechanical stiffness~\cite{Wang:2013,Wu:2016}. However, integration of these constraints into the VD computation is computationally expensive. Thus we handle them as a postprocess after we generate the support-free hollowed volume.

\paragraph{Design Variables}
We define a design variable $\gamma_E \in [0,1]$, as a shrinking factor, for each ellipse $E$ inside $\mathcal{M}$. A value $\gamma_E =0 $ means that $E$ is totally filled with solid.
The basic idea is that shrinking of ellipses can shift the center of gravity of the model and can improve its mechanical stiffness as more material is filled.
Thus we can easily formulate the optimization according to a specific objective function. In particular, we discuss about the optimization with respect to static balance as an example. The optimization for other constraints can be similarly achieved.
%
%

\paragraph{Static Balance}
An object is self-balanced when the vertical projection of its gravity center lies in the convex hull of its contact points with the ground.
As shown in Figure~\ref{fig:balance}, it is intuitive that shrinking ellipses on the left hand side of the gravity center will shift it leftwards, i.e., closer to the convex hull of its contact points.
It is easy to formulate an optimization of minimizing the horizontal distance between the gravity center and the boundary of the convex hull.
Our optimizer thus tries to reach a balance by shrinking some ellipses and thus shifting the gravity center into the convex hull.

\FigTwo{balance}{0.36}{0.36}
{
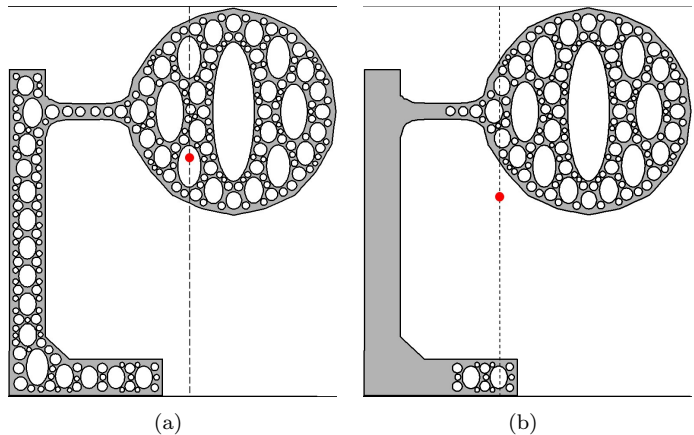
Optimization of static balance (2D case). (a) The hollowed object cannot stand by itself (left). (b) It is optimized to a self-balanced object using our optimizer (right). The red dots denote the gravity center.
}

\section{Experimental Results}

\paragraph{Computational Platform}
We have implemented our algorithm in C++ on a standard desktop PC with Intel(R) Core(Tm) i7-4790K CPU@4.0 GHz and 16 GB of RAM.
%
Thanks to the efficient implementation of TOI-EinP, the VD generation of polygons and ellipses is fast and the ellipse hollowing is less than 30 seconds for all examples.

\paragraph{Printer Configuration and Parameters}
We fabricate the objects using a commercial FDM 3D printer: The Ultimaker 2+ with tray size of
$223 \texttt{mm}\times 223\texttt{mm}\times 205\texttt{mm}$. The printable layer thickness of the printer (printing precision) ranges from $0.1 \texttt{mm}$ to $0.4 \texttt{mm}$ and we use a value of $\sigma_0 = 0.2\texttt{mm}$.
We test the plastic PLA material used in the 3D printing and set the maximally allowed overhang-angle as $\theta_0=60^{\circ}$  and the maximal length of printable horizontal hangover as $\delta_0=5 \texttt{mm}$.

\paragraph{Manufacture Setting}

%
After we generate the hollowed models, we add supporting structures for the exterior part of the models but do not add any interior support. After the models are fabricated, we manually remove the exterior supports.
All models have been successfully printed which reveals that the hollowed interior of the models is printed without any problem. We also validate this by printing and checking only half of the models  which will be shown later.

\paragraph{Experiments}

Figure~\ref{fig:hollowing-volume} shows an example of hollowed bunny model with different cross-sections.
Figure~\ref{fig:cross-sections-hollowing} shows a 3D hollowed volume of kitten model with a set of cross-sections, and then adding more in later passes.
Figure~\ref{fig:balance-3D} shows two hollowed hanging balls. The left one cannot stand by itself. After using our optimizer, it can be optimized to be well balanced in the right by filling material in the elliptic voids of the column.
Figure~\ref{fig:gallery} shows photos of the fabricated bunny model and kitten model which are hollowed by our method. The models are successfully fabricated without adding any extra support in the interior voids.

\FigOne{hollowing-volume}{1.2}{A hollowed bunny model. (a) the hollowed model; (b-d) different cross-sections.}



\FigFour{cross-sections-hollowing}{0.45}{0.45}{0.45}{0.45}
{
A hollowed bunny model. (a) the hollowed model; (b-d) different cross-sections.
}



\FigTwo{balance-3D}{0.48}{0.48}
{
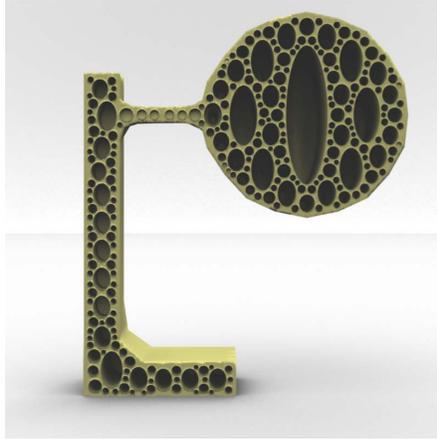
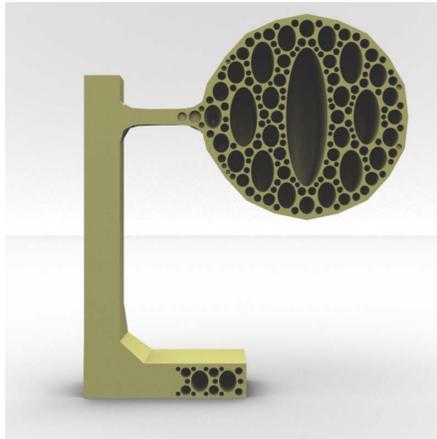
(a) The hollowed hanging ball may fall down without balance optimization (left) (b) while the optimized one is standing stable.
}



\FigOne{gallery}{1.0}
{
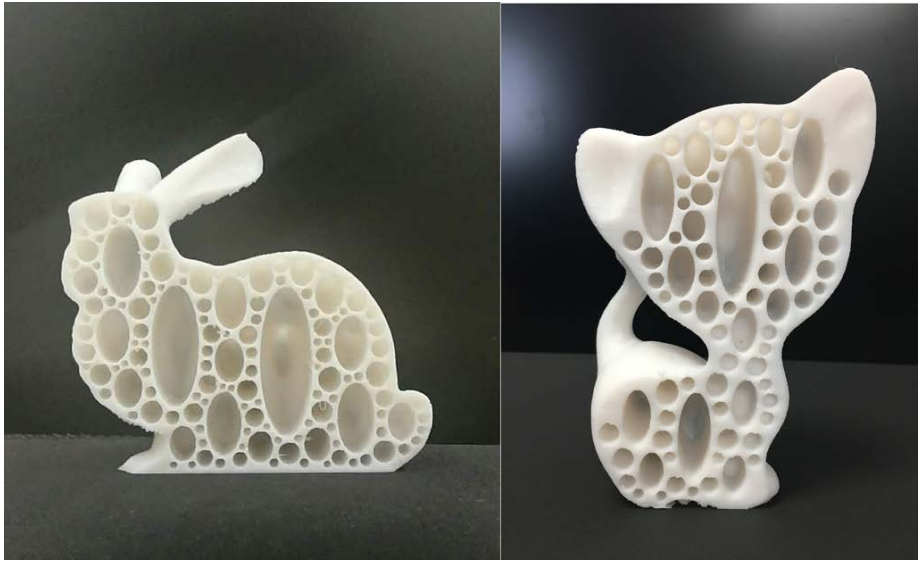
Photos of the fabricated bunny model and kitten model hollowed by our method.
}



\paragraph{Performance of the TOI-EinP Algorithm for Constructing VD}

We conducted computational efficiency test using two models: the bunny and the hanging ball models.
From the bunny model, we produced 121 planes resulting 219 polygons as some planes contain more than one polygon.
The smallest and the largest polygons have 8 and 655 P-edges, respectively.
For experimental purpose, we enforced to pack 100 ellipses into each polygon ignoring the mechanical and physicochemical constraints.
Figure \ref{fig:computation-time-profile}(a) shows computation time vs. polygon size in terms of the input P-edges.
The top-most black curve: the total time;
The next red one: the time for incrementing all the in-disks of the ellipses into the VD structure;
The next green one: that for finding the maximum clearance probe;
The blue one: that for constructing the VD of the at-disks and on-disks.
Note that the total time is weakly super-linear mainly due to the increment process of in-disks which is believed to be caused by the mapping mechanism of equivalent V-vertices between $\mathcal{VD}(\mathcal{D})$ and $\mathcal{VD}(\tilde{\mathcal{P}},\mathcal{E})$.
We used the map in the C++ template which is implemented by a binary search tree, taking $O(\log n)$ time for each query for $n$ entities.
With this model, we used non-uniform on-disk generation method using a bucket system for the acceleration.

From the hanging ball model, we produced 88 planes resulting 141 polygons: The smallest and the largest polygons have 25 and 232 P-edges, respectively.
As this model consists of several large planar faces together with smaller ones, the polygons have several long P-edges together with short ones.
This is common in many engineering products.
Hence, we subdivided each P-edge into a set of P-edges of the length defined by the previously stated rule.
Figure \ref{fig:computation-time-profile}(b) also shows computation time vs. polygon size in terms of the input P-edges.
Note that the correlation is very weak compared to the bunny model as is expected because of the big variation of the P-edge lengths.
Figure \ref{fig:computation-time-profile}(c) shows computation time vs. the number of subdivided P-edges:
The curves are fairly well correlated in a slightly super-linear fashion.
The big gab between the two clusters of data is due to many subdivided P-edges through very long input P-edges.
For example, the left-most data in the right cluster in Figure \ref{fig:computation-time-profile}(c) corresponds to a polygon with 234 input P-edges which was subdivided into 513 shorter P-edges.
Figure \ref{fig:computation-time-profile}(d) shows the number of subdivided P-edges vs. the number of input P-edges.
The bunny model is expectedly a straight line whereas the hanging ball model shows bumpy curve which is similar to the curves in Figure \ref{fig:computation-time-profile}(b).
The relationship between Figure \ref{fig:computation-time-profile}(b) and (c) can be explained by Figure \ref{fig:computation-time-profile}(d).
Figure \ref{fig:computation-time-profile}(e) shows the packing ratio of the 100 ellipses in each polygon.
The bunny model is expectedly smooth with decreasing pattern for bigger polygons as we incremented only 100 ellipses whereas the curve of the hanging ball model is bumpy.

\FigFive{computation-time-profile}{0.468}{0.468}{0.468}{0.468}{0.468}
{
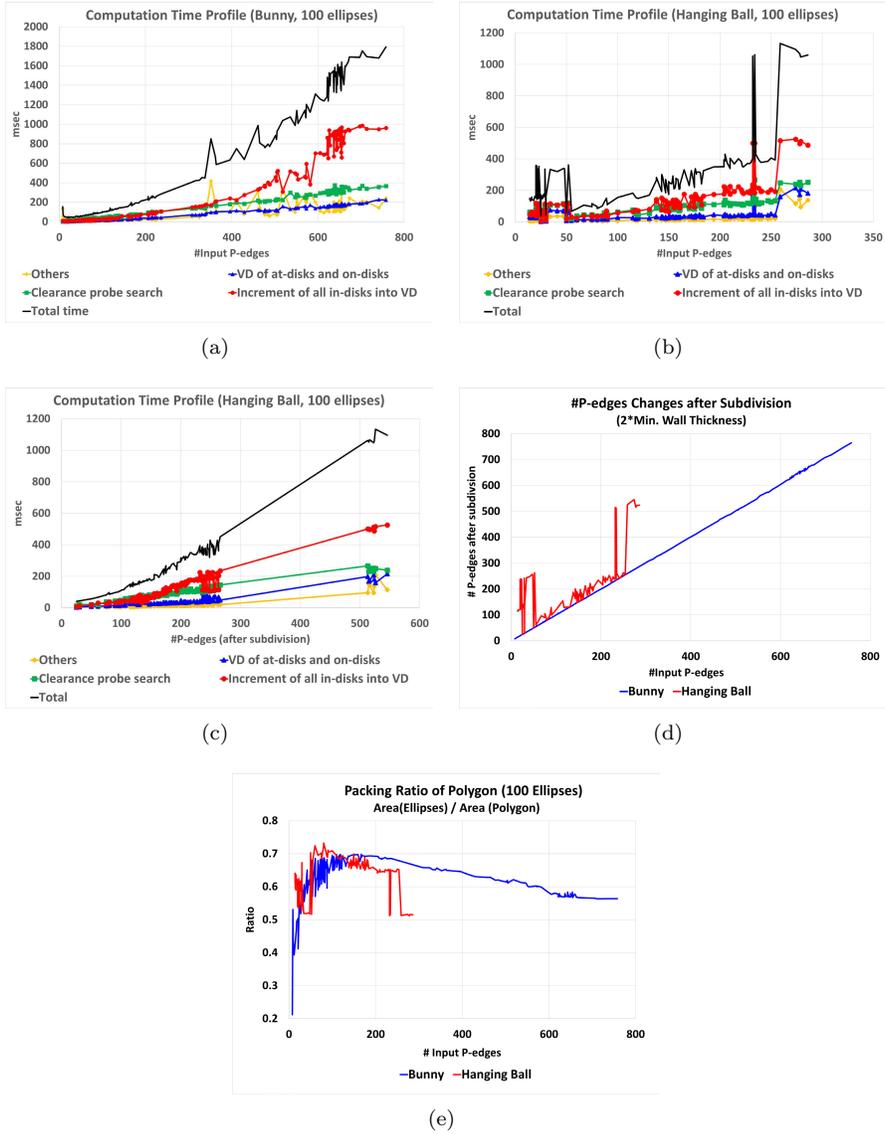
Computation time profile of the TOI-EinP algorithm.
(a) Bunny model (Time vs. \# input P-edges).
(b) Hanging ball model (Time vs. \# input P-edges).
(c) Hanging ball model (Time vs. \# subdivided P-edges).
(d) \# P-edges changes of both Bunny and Hanging ball models after subdivision.
(e) Packing ratio of polygons in both Bunny and Hanging ball models.
}

\paragraph{Comparison to rhombic cell structure}

The work of~\cite{Wu:2016} adopts rhombic cell structure, which have $C^0$ discontinuity on boundaries, to generate support-free interior voids for 3D shapes. We compare our method with this method as shown in Figure~\ref{fig:Comparison-P-Model}.
We apply two methods on the same P model with similar hollowing ratios. Then we fix the bottom of the model and conduct an identical external load on it, respectively. From the stress map we can see that the result generated by~\cite{Wu:2016} suffers the problem of stress concentration at the region marked in red, which generally happens in discontinuity.
This does not happen in our method.

\FigOne{Comparison-P-Model}{0.85}
{
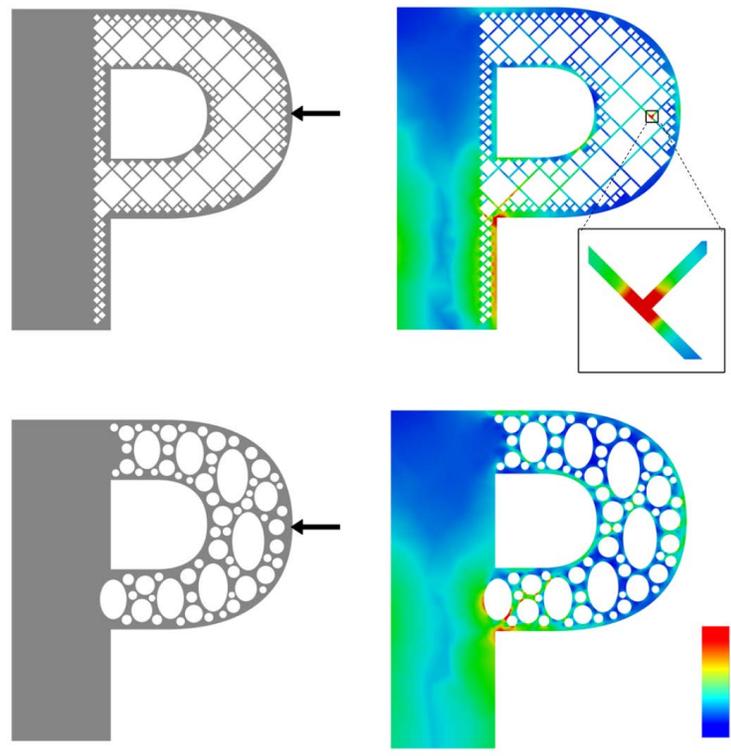
Comparison with [Wu et al. 2016b]. Upper row: the hollowed P model using [Wu et al. 2016b] with a hollowing ratio of 24.3\%; Lower row: the hollowed P model using our method with a hollowing ratio of 25.7\%. The bottom of the model is fixed and one identical external load is conducted on the right, respectively, as shown by the arrow. The right figures show the color maps of stress. It is seen that the region marked in red in the upper-right figure suffers the problem of stress concentration, i.e., the stress there is very high.
}


\section{Conclusions and Future Work}

In this paper we propose a novel approach for generating support-free interior hollowing for general 3D shapes. The generated 3D shapes can be directly fabricated with FDM 3D printers without any usage of extra supports in interior voids. This is based on the observation of a family of support-free ellipses and is achieved by hollowing 2D shapes with these ellipses. Then the interior ellipses are extruded into volume for generating hollowed 3D shapes.
We also develop a new, efficient and robust algorithm for the Voronoi diagram polygons and the first algorithm for the Voronoi diagram of ellipses within a polygon, both based on the topology-oriented incremental approach, which are quite useful for generating the ellipse packing in 2D shapes.
With the sizes of ellipses as design variables, the optimization according to a specific objective function, e.g., static stability, can be easily formulated.
Experimental results have shown the practicability and feasibility of our proposed approach.

\paragraph{Limitation and Future work}
Our research opens many directions for future studies.
First, the packing results can be further optimized by optimizing the positions and sizes of the ellipses for the purpose of increasing packing ratio.
Second, it is expected to extend our approach for generating support-free ellipsoids for 3D shapes. This is feasible but needs more effort.
Last but not the least, we are interested in studying general support-free shapes for additive manufacturing, which is a promising direction for geometric modeling and processing.

\section*{Acknowledgements}
This work was supported by the National Research Foundation of Korea(NRF) grant funded by the Korea government(MSIP) (No. 2017R1A3B1023591).

\section*{References}

\bibliography{EllipseHollowing}

\begin{thebibliography}{54}
\providecommand{\natexlab}[1]{#1}
\providecommand{\url}[1]{\texttt{#1}}
\providecommand{\href}[2]{#2}
\providecommand{\path}[1]{#1}
\providecommand{\eprint}[1]{\href{http://arxiv.org/abs/#1}{\path{#1}}}
\providecommand{\DOIprefix}{doi:}
\providecommand{\ArXivprefix}{arXiv:}
\providecommand{\URLprefix}{URL: }
\providecommand{\Pubmedprefix}{pmid:}
\providecommand{\doi}[1]{\href{http://dx.doi.org/#1}{\path{#1}}}
\providecommand{\Pubmed}[1]{\href{pmid:#1}{\path{#1}}}
\providecommand{\BIBand}{and}
\providecommand{\bibinfo}[2]{#2}
\ifx\xfnm\undefined \def\xfnm[#1]{\unskip,\space#1}\fi
\bibitem[{Stava et~al.(2012)Stava, Vanek, Benes, Carr and
  M\v{e}ch}]{Stava:2012}
\bibinfo{author}{Stava\xfnm[ O.]}, \bibinfo{author}{Vanek\xfnm[ J.]},
  \bibinfo{author}{Benes\xfnm[ B.]}, \bibinfo{author}{Carr\xfnm[ N.]},
  \bibinfo{author}{M\v{e}ch\xfnm[ R.]}.
\newblock \bibinfo{title}{Stress relief: Improving structural strength of {3D}
  printable objects}.
\newblock \bibinfo{journal}{ACM Trans Graph}
  \bibinfo{year}{2012};\bibinfo{volume}{31}(\bibinfo{number}{4}):\bibinfo{pages}{48:1--48:11}.
\bibitem[{Wang et~al.(2013)Wang, Wang, Yang, Liu, Tong, Tong
  et~al.}]{Wang:2013}
\bibinfo{author}{Wang\xfnm[ W.]}, \bibinfo{author}{Wang\xfnm[ T.Y.]},
  \bibinfo{author}{Yang\xfnm[ Z.]}, \bibinfo{author}{Liu\xfnm[ L.]},
  \bibinfo{author}{Tong\xfnm[ X.]}, \bibinfo{author}{Tong\xfnm[ W.]}, et~al.
\newblock \bibinfo{title}{Cost-effective printing of {3D} objects with
  skin-frame structures}.
\newblock \bibinfo{journal}{ACM Trans Graph}
  \bibinfo{year}{2013};\bibinfo{volume}{32}(\bibinfo{number}{6}):\bibinfo{pages}{177:1--177:10}.
\bibitem[{Lu et~al.(2014)Lu, Sharf, Zhao, Wei, Fan, Chen et~al.}]{Lu:2014}
\bibinfo{author}{Lu\xfnm[ L.]}, \bibinfo{author}{Sharf\xfnm[ A.]},
  \bibinfo{author}{Zhao\xfnm[ H.]}, \bibinfo{author}{Wei\xfnm[ Y.]},
  \bibinfo{author}{Fan\xfnm[ Q.]}, \bibinfo{author}{Chen\xfnm[ X.]}, et~al.
\newblock \bibinfo{title}{Build-to-last: Strength to weight {3D} printed
  objects}.
\newblock \bibinfo{journal}{ACM Trans Graph}
  \bibinfo{year}{2014};\bibinfo{volume}{33}(\bibinfo{number}{4}):\bibinfo{pages}{97:1--97:10}.
\bibitem[{Strano et~al.(2013)Strano, Hao, Everson and Evans}]{Strano:2013}
\bibinfo{author}{Strano\xfnm[ G.]}, \bibinfo{author}{Hao\xfnm[ L.]},
  \bibinfo{author}{Everson\xfnm[ R.M.]}, \bibinfo{author}{Evans\xfnm[ K.E.]}.
\newblock \bibinfo{title}{A new approach to the design and optimisation of
  support structures in additive manufacturing}.
\newblock \bibinfo{journal}{International Journal of Advanced Manufacturing
  Technology}
  \bibinfo{year}{2013};\bibinfo{volume}{66}(\bibinfo{number}{9}):\bibinfo{pages}{1247--1254}.
\bibitem[{Dumas et~al.(2014)Dumas, Hergel and Lefebvre}]{Dumas:2014}
\bibinfo{author}{Dumas\xfnm[ J.]}, \bibinfo{author}{Hergel\xfnm[ J.]},
  \bibinfo{author}{Lefebvre\xfnm[ S.]}.
\newblock \bibinfo{title}{Bridging the gap: Automated steady scaffoldings for
  {3D} printing}.
\newblock \bibinfo{journal}{ACM Transactions on Graphics}
  \bibinfo{year}{2014};\bibinfo{volume}{33}(\bibinfo{number}{4}):\bibinfo{pages}{1--10}.
\bibitem[{Vanek et~al.(2014)Vanek, Galicia and Benes}]{Vanek:2014}
\bibinfo{author}{Vanek\xfnm[ J.]}, \bibinfo{author}{Galicia\xfnm[ J.A.G.]},
  \bibinfo{author}{Benes\xfnm[ B.]}.
\newblock \bibinfo{title}{Clever support: Efficient support structure
  generation for digital fabrication}.
\newblock \bibinfo{journal}{Comput Graph Forum}
  \bibinfo{year}{2014};\bibinfo{volume}{33}(\bibinfo{number}{5}):\bibinfo{pages}{117--125}.
\bibitem[{Langelaar(2016)}]{Langelaar:2016}
\bibinfo{author}{Langelaar\xfnm[ M.]}.
\newblock \bibinfo{title}{Topology optimization of 3d self-supporting
  structures for additive manufacturing}.
\newblock \bibinfo{journal}{Additive Manufacturing}
  \bibinfo{year}{2016};\bibinfo{volume}{12;Part A}:\bibinfo{pages}{60--70}.
\bibitem[{Wu et~al.(2016{\natexlab{a}})Wu, Wang, Zhang and
  Westermann}]{Wu:2016}
\bibinfo{author}{Wu\xfnm[ J.]}, \bibinfo{author}{Wang\xfnm[ C.C.L.]},
  \bibinfo{author}{Zhang\xfnm[ X.]}, \bibinfo{author}{Westermann\xfnm[ R.]}.
\newblock \bibinfo{title}{Self-supporting rhombic infill structures for
  additive manufacturing}.
\newblock \bibinfo{journal}{Computer-Aided Design}
  \bibinfo{year}{2016}{\natexlab{a}};\bibinfo{volume}{80}:\bibinfo{pages}{32--42}.
\bibitem[{Reiner and Lefebvre(2016)}]{Reiner:2016}
\bibinfo{author}{Reiner\xfnm[ T.]}, \bibinfo{author}{Lefebvre\xfnm[ S.]}.
\newblock \bibinfo{title}{Interactive modeling of support-free shapes for
  fabrication}.
\newblock In: \bibinfo{booktitle}{EUROGRAPHICS}. \bibinfo{year}{2016},.
\bibitem[{Pilkey and Pilkey(2008)}]{Walter:2008}
\bibinfo{author}{Pilkey\xfnm[ W.D.]}, \bibinfo{author}{Pilkey\xfnm[ D.F.]}.
\newblock \bibinfo{title}{Peterson's stress concentration factors}.
\newblock \bibinfo{edition}{3} ed.; \bibinfo{publisher}{Wiley};
  \bibinfo{year}{2008}.
\newblock ISBN \bibinfo{isbn}{978-0-470-04824-5}.
\bibitem[{Karthikeyan(2016)}]{GeekStress:2016}
\bibinfo{author}{Karthikeyan\xfnm[ K.]}.
\newblock \bibinfo{title}{Why hatches and doorways in ships and airplanes are
  oval?}
\newblock \bibinfo{year}{2016}.
\newblock \bibinfo{note}{\url{
  https://geekswipe.net/science/physics/why-hatches-and-doorways-in-ships-and-airplanes-are-oval/
  }}.
\bibitem[{Zhang et~al.(2015)Zhang, Xia, Wang, Yang, Tu and Wang}]{Zhang:2015}
\bibinfo{author}{Zhang\xfnm[ X.]}, \bibinfo{author}{Xia\xfnm[ Y.]},
  \bibinfo{author}{Wang\xfnm[ J.]}, \bibinfo{author}{Yang\xfnm[ Z.]},
  \bibinfo{author}{Tu\xfnm[ C.]}, \bibinfo{author}{Wang\xfnm[ W.]}.
\newblock \bibinfo{title}{Medial axis tree-an internal supporting structure for
  {3D} printing}.
\newblock \bibinfo{journal}{Computer Aided Geometric Design}
  \bibinfo{year}{2015};\bibinfo{volume}{35-36}(\bibinfo{number}{5}):\bibinfo{pages}{149--162}.
\bibitem[{Hu et~al.(2015)Hu, Jin and Wang}]{Hu:2015}
\bibinfo{author}{Hu\xfnm[ K.]}, \bibinfo{author}{Jin\xfnm[ S.]},
  \bibinfo{author}{Wang\xfnm[ C.C.]}.
\newblock \bibinfo{title}{Support slimming for single material based additive
  manufacturing}.
\newblock \bibinfo{journal}{Computer-Aided Design}
  \bibinfo{year}{2015};\bibinfo{volume}{65}:\bibinfo{pages}{1--10}.
\bibitem[{Musialski et~al.(2015)Musialski, Auzinger, Birsak, Wimmer and
  Kobbelt}]{Musialski:2015}
\bibinfo{author}{Musialski\xfnm[ P.]}, \bibinfo{author}{Auzinger\xfnm[ T.]},
  \bibinfo{author}{Birsak\xfnm[ M.]}, \bibinfo{author}{Wimmer\xfnm[ M.]},
  \bibinfo{author}{Kobbelt\xfnm[ L.]}.
\newblock \bibinfo{title}{Reduced-order shape optimization using offset
  surfaces}.
\newblock \bibinfo{journal}{ACM Transactions on Graphics}
  \bibinfo{year}{2015};\bibinfo{volume}{34}(\bibinfo{number}{4}):\bibinfo{pages}{102;1--9}.
\bibitem[{Pr{\'e}vost et~al.(2013)Pr{\'e}vost, Whiting, Lefebvre and
  Sorkine-Hornung}]{Prevost:2013}
\bibinfo{author}{Pr{\'e}vost\xfnm[ R.]}, \bibinfo{author}{Whiting\xfnm[ E.]},
  \bibinfo{author}{Lefebvre\xfnm[ S.]}, \bibinfo{author}{Sorkine-Hornung\xfnm[
  O.]}.
\newblock \bibinfo{title}{Make it stand: Balancing shapes for {3D}
  fabrication}.
\newblock \bibinfo{journal}{ACM Trans Graph}
  \bibinfo{year}{2013};\bibinfo{volume}{32}(\bibinfo{number}{4}).
\bibitem[{Christiansen et~al.(2015)Christiansen, Schmidt and
  B{\ae}rentzen}]{Christiansen:2015}
\bibinfo{author}{Christiansen\xfnm[ A.N.]}, \bibinfo{author}{Schmidt\xfnm[
  R.]}, \bibinfo{author}{B{\ae}rentzen\xfnm[ J.A.]}.
\newblock \bibinfo{title}{Automatic balancing of 3d models}.
\newblock \bibinfo{journal}{Computer-Aided Design}
  \bibinfo{year}{2015};\bibinfo{volume}{58}:\bibinfo{pages}{236--241}.
\bibitem[{B\"{a}cher et~al.(2014)B\"{a}cher, Whiting, Bickel and
  Sorkine-Hornung}]{Bacher:2014}
\bibinfo{author}{B\"{a}cher\xfnm[ M.]}, \bibinfo{author}{Whiting\xfnm[ E.]},
  \bibinfo{author}{Bickel\xfnm[ B.]}, \bibinfo{author}{Sorkine-Hornung\xfnm[
  O.]}.
\newblock \bibinfo{title}{Spin-it: Optimizing moment of inertia for spinnable
  objects}.
\newblock \bibinfo{journal}{ACM Trans Graph}
  \bibinfo{year}{2014};\bibinfo{volume}{33}(\bibinfo{number}{4}):\bibinfo{pages}{96:1--96:10}.
\bibitem[{Deaton and Grandhi(2014)}]{Deaton:2014}
\bibinfo{author}{Deaton\xfnm[ J.D.]}, \bibinfo{author}{Grandhi\xfnm[ R.V.]}.
\newblock \bibinfo{title}{A survey of structural and multidisciplinary
  continuum topology optimization: post 2000}.
\newblock \bibinfo{journal}{Structural and Multidisciplinary Optimization}
  \bibinfo{year}{2014};\bibinfo{volume}{49}(\bibinfo{number}{1}):\bibinfo{pages}{1--38}.
\bibitem[{Wu et~al.(2016{\natexlab{b}})Wu, Dick and Westermann}]{Wu:2016-TVCG}
\bibinfo{author}{Wu\xfnm[ J.]}, \bibinfo{author}{Dick\xfnm[ C.]},
  \bibinfo{author}{Westermann\xfnm[ R.]}.
\newblock \bibinfo{title}{A system for high-resolution topology optimization}.
\newblock \bibinfo{journal}{IEEE Transactions on Visualization and Computer
  Graphics}
  \bibinfo{year}{2016}{\natexlab{b}};\bibinfo{volume}{22}(\bibinfo{number}{3}):\bibinfo{pages}{1195--1208}.
\bibitem[{Hu et~al.(2014)Hu, Li, Zhang and Cohen-Or}]{Hu:2014-Pyramidal}
\bibinfo{author}{Hu\xfnm[ R.]}, \bibinfo{author}{Li\xfnm[ H.]},
  \bibinfo{author}{Zhang\xfnm[ H.]}, \bibinfo{author}{Cohen-Or\xfnm[ D.]}.
\newblock \bibinfo{title}{Approximate pyramidal shape decomposition}.
\newblock \bibinfo{journal}{ACM Trans Graph}
  \bibinfo{year}{2014};\bibinfo{volume}{33}(\bibinfo{number}{6}):\bibinfo{pages}{213:1--10}.
\bibitem[{Buchalter and Bradley(1992)}]{BuchalterBradley92}
\bibinfo{author}{Buchalter\xfnm[ B.J.]}, \bibinfo{author}{Bradley\xfnm[ R.M.]}.
\newblock \bibinfo{title}{Orientational order in amorphous packings of
  ellipses}.
\newblock \bibinfo{journal}{Journal of Physics A: Mathematical and General}
  \bibinfo{year}{1992};\bibinfo{volume}{26}(\bibinfo{number}{3}):\bibinfo{pages}{L1219--L1224}.
\bibitem[{Schreck et~al.(2010)Schreck, Xu and O'Hern}]{SchreckEtal10}
\bibinfo{author}{Schreck\xfnm[ C.F.]}, \bibinfo{author}{Xu\xfnm[ N.]},
  \bibinfo{author}{O'Hern\xfnm[ C.S.]}.
\newblock \bibinfo{title}{A comparison of jamming behavior in systems composed
  of dimer- and ellipse-shaped particles}.
\newblock \bibinfo{journal}{Soft Matter}
  \bibinfo{year}{2010};\bibinfo{volume}{6}(\bibinfo{number}{13}):\bibinfo{pages}{2960--2969}.
\bibitem[{Donev et~al.(2004)Donev, Torquato, Stillinger and
  Connelly}]{DonevEtal04}
\bibinfo{author}{Donev\xfnm[ A.]}, \bibinfo{author}{Torquato\xfnm[ S.]},
  \bibinfo{author}{Stillinger\xfnm[ F.H.]}, \bibinfo{author}{Connelly\xfnm[
  R.]}.
\newblock \bibinfo{title}{Jamming in hard sphere and disk packings}.
\newblock \bibinfo{journal}{Journal of Applied Physics}
  \bibinfo{year}{2004};\bibinfo{volume}{95}(\bibinfo{number}{3}):\bibinfo{pages}{989--999}.
\bibitem[{Donev et~al.(2007)Donev, Connelly, Stillinger and
  Torquato}]{DonevEtal07}
\bibinfo{author}{Donev\xfnm[ A.]}, \bibinfo{author}{Connelly\xfnm[ R.]},
  \bibinfo{author}{Stillinger\xfnm[ F.H.]}, \bibinfo{author}{Torquato\xfnm[
  S.]}.
\newblock \bibinfo{title}{Underconstrained jammed packings of nonspherical hard
  particles: Ellipses and ellipsoids}.
\newblock \bibinfo{journal}{Physical Review E}
  \bibinfo{year}{2007};\bibinfo{volume}{75}(\bibinfo{number}{5}):\bibinfo{pages}{051304}.
\bibitem[{Hit(2013)}]{Hitti:2013}
\bibinfo{title}{Optimized dropping and rolling (odr) method for packing of
  poly-disperse spheres}.
\newblock \bibinfo{journal}{Applied Mathematical Modelling}
  \bibinfo{year}{2013};\bibinfo{volume}{37}(\bibinfo{number}{8}):\bibinfo{pages}{5715--5722}.
\bibitem[{Birgin et~al.(2016)Birgin, Lobato and Mart\'{\i}nez}]{Birgin:2016}
\bibinfo{author}{Birgin\xfnm[ E.G.]}, \bibinfo{author}{Lobato\xfnm[ R.D.]},
  \bibinfo{author}{Mart\'{\i}nez\xfnm[ J.M.]}.
\newblock \bibinfo{title}{Packing ellipsoids by nonlinear optimization}.
\newblock \bibinfo{journal}{Journal of Global Optimization}
  \bibinfo{year}{2016};\bibinfo{volume}{65}(\bibinfo{number}{4}):\bibinfo{pages}{709--743}.
\bibitem[{Delaney et~al.(2005)Delaney, Weaire, Hutzler and
  Merphy}]{DelaneyEtal05}
\bibinfo{author}{Delaney\xfnm[ G.]}, \bibinfo{author}{Weaire\xfnm[ D.]},
  \bibinfo{author}{Hutzler\xfnm[ S.]}, \bibinfo{author}{Merphy\xfnm[ S.]}.
\newblock \bibinfo{title}{Random packing of elliptical disks}.
\newblock \bibinfo{journal}{Philisophical Magazine Letters}
  \bibinfo{year}{2005};\bibinfo{volume}{85}(\bibinfo{number}{2}):\bibinfo{pages}{89--96}.
\bibitem[{Lozano et~al.(2016)Lozano, Roehl, Celes and Gattass}]{Lozano:2016}
\bibinfo{author}{Lozano\xfnm[ E.]}, \bibinfo{author}{Roehl\xfnm[ D.]},
  \bibinfo{author}{Celes\xfnm[ W.]}, \bibinfo{author}{Gattass\xfnm[ M.]}.
\newblock \bibinfo{title}{An efficient algorithm to generate random sphere
  packs in arbitrary domains}.
\newblock \bibinfo{journal}{Comput Math Appl}
  \bibinfo{year}{2016};\bibinfo{volume}{71}(\bibinfo{number}{8}):\bibinfo{pages}{1586--1601}.
\bibitem[{Jodrey and Tory(1985)}]{Jodrey:1985}
\bibinfo{author}{Jodrey\xfnm[ W.S.]}, \bibinfo{author}{Tory\xfnm[ E.M.]}.
\newblock \bibinfo{title}{Computer simulation of close random packing of equal
  spheres}.
\newblock \bibinfo{journal}{Phys Rev A}
  \bibinfo{year}{1985};\bibinfo{volume}{32}:\bibinfo{pages}{2347--2351}.
\bibitem[{Lubachevsky and Stillinger(1990)}]{Lubachevsky:1990}
\bibinfo{author}{Lubachevsky\xfnm[ B.D.]}, \bibinfo{author}{Stillinger\xfnm[
  F.H.]}.
\newblock \bibinfo{title}{Geometric properties of random disk packings}.
\newblock \bibinfo{journal}{Journal of Statistical Physics}
  \bibinfo{year}{1990};\bibinfo{volume}{60}(\bibinfo{number}{5}):\bibinfo{pages}{561--583}.
\bibitem[{Lubachevsky(1991)}]{Lubachevsky91}
\bibinfo{author}{Lubachevsky\xfnm[ B.D.]}.
\newblock \bibinfo{title}{How to simulate billiards and similar systems}.
\newblock \bibinfo{journal}{Journal of Computational Physics}
  \bibinfo{year}{1991};\bibinfo{volume}{94}(\bibinfo{number}{2}):\bibinfo{pages}{255--283}.
\bibitem[{Specht(2015)}]{Specht15}
\bibinfo{author}{Specht\xfnm[ E.]}.
\newblock \bibinfo{title}{A precise algorithm to detect voids in polydisperse
  circle packings}.
\newblock In: \bibinfo{booktitle}{Proc. R. Soc. A}. \bibinfo{year}{2015}, p.
  \bibinfo{pages}{20150421}.
\bibitem[{Sugihara et~al.(2004)Sugihara, Sawai, Sano, Kim and
  Kim}]{SugiharaEtal04}
\bibinfo{author}{Sugihara\xfnm[ K.]}, \bibinfo{author}{Sawai\xfnm[ M.]},
  \bibinfo{author}{Sano\xfnm[ H.]}, \bibinfo{author}{Kim\xfnm[ D.S.]},
  \bibinfo{author}{Kim\xfnm[ D.]}.
\newblock \bibinfo{title}{Disk packing for the estimation of the size of a wire
  bundle}.
\newblock \bibinfo{journal}{Japan Journal of Industrial and Applied
  Mathematics}
  \bibinfo{year}{2004};\bibinfo{volume}{21}(\bibinfo{number}{3}):\bibinfo{pages}{259--278}.
\bibitem[{Lee et~al.(2016)Lee, Sugihara and Kim}]{LmwEtal16}
\bibinfo{author}{Lee\xfnm[ M.]}, \bibinfo{author}{Sugihara\xfnm[ K.]},
  \bibinfo{author}{Kim\xfnm[ D.S.]}.
\newblock \bibinfo{title}{Topology-oriented incremental algorithm for the
  robust construction of the voronoi diagrams of disks}.
\newblock \bibinfo{journal}{ACM Transactions on Mathematical Software}
  \bibinfo{year}{2016};\bibinfo{volume}{43}(\bibinfo{number}{2}):\bibinfo{pages}{14:1--14:23}.
\bibitem[{Okabe et~al.(1999)Okabe, Boots, Sugihara and Chiu}]{OkabeEtal99}
\bibinfo{author}{Okabe\xfnm[ A.]}, \bibinfo{author}{Boots\xfnm[ B.]},
  \bibinfo{author}{Sugihara\xfnm[ K.]}, \bibinfo{author}{Chiu\xfnm[ S.N.]}.
\newblock \bibinfo{title}{Spatial Tessellations: Concepts and Applications of
  {V}oronoi Diagrams}.
\newblock \bibinfo{edition}{2nd} ed.; \bibinfo{address}{Chichester}:
  \bibinfo{publisher}{John Wiley \& Sons}; \bibinfo{year}{1999}.
\bibitem[{Kim et~al.(1995)Kim, Hwang and Park}]{KdsEtal95}
\bibinfo{author}{Kim\xfnm[ D.S.]}, \bibinfo{author}{Hwang\xfnm[ I.K.]},
  \bibinfo{author}{Park\xfnm[ B.J.]}.
\newblock \bibinfo{title}{Representing the {V}oronoi diagram of a simple
  polygon using rational quadratic {B}$\acute{e}$zier curves}.
\newblock \bibinfo{journal}{Computer-Aided Design}
  \bibinfo{year}{1995};\bibinfo{volume}{27}(\bibinfo{number}{8}):\bibinfo{pages}{605--614}.
\bibitem[{Sugihara and Iri(1992)}]{SugiharaIri92}
\bibinfo{author}{Sugihara\xfnm[ K.]}, \bibinfo{author}{Iri\xfnm[ M.]}.
\newblock \bibinfo{title}{Construction of the {V}oronoi diagram for ``one
  million'' generators in single-precision arithmetic}.
\newblock \bibinfo{journal}{Proceedings of the IEEE}
  \bibinfo{year}{1992};\bibinfo{volume}{80}(\bibinfo{number}{9}):\bibinfo{pages}{1471--1484}.
\bibitem[{Held(2001)}]{Held01}
\bibinfo{author}{Held\xfnm[ M.]}.
\newblock \bibinfo{title}{{VRONI}: An engineering approach to the reliable and
  efficient computation of {V}oronoi diagrams of points and line segments}.
\newblock \bibinfo{journal}{Computational Geometry - Theory and Applications}
  \bibinfo{year}{2001};\bibinfo{volume}{18}(\bibinfo{number}{2}):\bibinfo{pages}{95--123}.
\bibitem[{Held and Huber(2009)}]{HeldHuber09}
\bibinfo{author}{Held\xfnm[ M.]}, \bibinfo{author}{Huber\xfnm[ S.]}.
\newblock \bibinfo{title}{Topology-oriented incremental computation of
  {V}oronoi diagrams of circular arcs and straight-line segments}.
\newblock \bibinfo{journal}{Computer-Aided Design}
  \bibinfo{year}{2009};\bibinfo{volume}{41}(\bibinfo{number}{5}):\bibinfo{pages}{327--338}.
\bibitem[{CGAL(2016)}]{CGALHome}
\bibinfo{author}{CGAL\xfnm[]}.
\newblock \bibinfo{title}{{CGAL} {L}ibrary {H}omepage}.
\newblock \bibinfo{year}{2016}.
\newblock \bibinfo{note}{{\tt http://www.cgal.org/}}.
\bibitem[{Alliez et~al.(2007)Alliez, Delage, Karavelas, Pion, Teillaud and
  Yvinec}]{AlliezEtal07}
\bibinfo{author}{Alliez\xfnm[ P.]}, \bibinfo{author}{Delage\xfnm[ C.]},
  \bibinfo{author}{Karavelas\xfnm[ M.I.]}, \bibinfo{author}{Pion\xfnm[ S.]},
  \bibinfo{author}{Teillaud\xfnm[ M.]}, \bibinfo{author}{Yvinec\xfnm[ M.]}.
\newblock \bibinfo{title}{Delaunay tessellations and {V}oronoi diagrams in
  {CGAL} (draft)}.
\newblock In: \bibinfo{editor}{Rien van~de Weijgaert Gert~Vegter\xfnm[ J.R.]},
  \bibinfo{editor}{Icke\xfnm[ V.]}, editors. \bibinfo{booktitle}{Tessellations
  in the Sciences. Virtues, Techniques and Applications of Geometric Tilings}.
  \bibinfo{year}{2007},.
\bibitem[{Kim et~al.(2001{\natexlab{a}})Kim, Kim and Sugihara}]{KdsEtal01}
\bibinfo{author}{Kim\xfnm[ D.S.]}, \bibinfo{author}{Kim\xfnm[ D.]},
  \bibinfo{author}{Sugihara\xfnm[ K.]}.
\newblock \bibinfo{title}{{V}oronoi diagram of a circle set from {V}oronoi
  diagram of a point set: {I}. topology}.
\newblock \bibinfo{journal}{Computer Aided Geometric Design}
  \bibinfo{year}{2001}{\natexlab{a}};\bibinfo{volume}{18}:\bibinfo{pages}{541--562}.
\bibitem[{Kim et~al.(2001{\natexlab{b}})Kim, Kim and Sugihara}]{KdsEtal01b}
\bibinfo{author}{Kim\xfnm[ D.S.]}, \bibinfo{author}{Kim\xfnm[ D.]},
  \bibinfo{author}{Sugihara\xfnm[ K.]}.
\newblock \bibinfo{title}{Voronoi diagram of a circle set from {V}oronoi
  diagram of a point set: {II}. geometry}.
\newblock \bibinfo{journal}{Computer Aided Geometric Design}
  \bibinfo{year}{2001}{\natexlab{b}};\bibinfo{volume}{18}:\bibinfo{pages}{563--585}.
\bibitem[{Sugihara(1993)}]{Sugihara93}
\bibinfo{author}{Sugihara\xfnm[ K.]}.
\newblock \bibinfo{title}{Approximation of generalized voronoi diagrams by
  ordinary voronoi diagrams}.
\newblock \bibinfo{journal}{Graphical Models and Image Processing}
  \bibinfo{year}{1993};\bibinfo{volume}{55}(\bibinfo{number}{6}):\bibinfo{pages}{522--531}.
\bibitem[{Emiris et~al.(2013)Emiris, Tsigaridas and Tzoumas}]{EmirisEtal13}
\bibinfo{author}{Emiris\xfnm[ I.Z.]}, \bibinfo{author}{Tsigaridas\xfnm[ E.P.]},
  \bibinfo{author}{Tzoumas\xfnm[ G.M.]}.
\newblock \bibinfo{title}{Exact voronoi diagram of smooth convex
  pseudo-circles: General predicates, and implementation for ellipses}.
\newblock \bibinfo{journal}{Computer Aided Geometric Design}
  \bibinfo{year}{2013};\bibinfo{volume}{30}(\bibinfo{number}{8}):\bibinfo{pages}{760--777}.
\bibitem[{Emiris and Tzoumas(2005)}]{EmirisTzoumas05}
\bibinfo{author}{Emiris\xfnm[ I.Z.]}, \bibinfo{author}{Tzoumas\xfnm[ G.M.]}.
\newblock \bibinfo{title}{Algebraic study of the apollonius circle of three
  ellipses}.
\newblock In: \bibinfo{booktitle}{EuroCG}. \bibinfo{year}{2005}, p.
  \bibinfo{pages}{147--150}.
\bibitem[{Farouki and Johnstone(1994)}]{FaroukiJohnstone94}
\bibinfo{author}{Farouki\xfnm[ R.T.]}, \bibinfo{author}{Johnstone\xfnm[ J.K.]}.
\newblock \bibinfo{title}{The bisector of a point and a plane parametric
  curve}.
\newblock \bibinfo{journal}{Computer Aided Geometric Design}
  \bibinfo{year}{1994};\bibinfo{volume}{11}:\bibinfo{pages}{117--151}.
\bibitem[{Alt et~al.(2005)Alt, Cheong and Vigneron}]{AltEtal05}
\bibinfo{author}{Alt\xfnm[ H.]}, \bibinfo{author}{Cheong\xfnm[ O.]},
  \bibinfo{author}{Vigneron\xfnm[ A.]}.
\newblock \bibinfo{title}{The voronoi diagram of curved objects}.
\newblock \bibinfo{journal}{Discrete Comput Geom}
  \bibinfo{year}{2005};\bibinfo{volume}{34}:\bibinfo{pages}{439--453}.
\bibitem[{Farouki and Johnstone(1992)}]{FaroukiJohnstone94b}
\bibinfo{author}{Farouki\xfnm[ R.T.]}, \bibinfo{author}{Johnstone\xfnm[ J.K.]}.
\newblock \bibinfo{title}{Computing point/curve and curve/curve bisectors}.
\newblock In: \bibinfo{booktitle}{IMA Conference on the Mathematics of
  Surfaces}. \bibinfo{year}{1992},.
\bibitem[{Farouki and Ramamurthy(1998)}]{FaroukiRamamurthy98}
\bibinfo{author}{Farouki\xfnm[ R.T.]}, \bibinfo{author}{Ramamurthy\xfnm[ R.]}.
\newblock \bibinfo{title}{Degenerate point/curve and curve/curve bisectors
  arising in medial axis computations for planar domains with curved
  boundaries}.
\newblock \bibinfo{journal}{Computer Aided Geometric Design}
  \bibinfo{year}{1998};\bibinfo{volume}{15}(\bibinfo{number}{6}):\bibinfo{pages}{615--635}.
\bibitem[{Omirou and Demosthenous(2006)}]{OmirouDemosthenous06}
\bibinfo{author}{Omirou\xfnm[ S.L.]}, \bibinfo{author}{Demosthenous\xfnm[
  G.A.]}.
\newblock \bibinfo{title}{A new interpolation algorithm for tracing planar
  equidistant curves}.
\newblock \bibinfo{journal}{Robotics and Computer-Integrated Manufacturing}
  \bibinfo{year}{2006};\bibinfo{volume}{22}:\bibinfo{pages}{306--314}.
\bibitem[{Cao and Liu(2008)}]{CaoLiu08}
\bibinfo{author}{Cao\xfnm[ L.]}, \bibinfo{author}{Liu\xfnm[ J.]}.
\newblock \bibinfo{title}{Computation of medial axis and offset curves of
  curved boundaries in planar domain}.
\newblock \bibinfo{journal}{Computer-Aided Design}
  \bibinfo{year}{2008};\bibinfo{volume}{40}(\bibinfo{number}{4}):\bibinfo{pages}{465--475}.
\bibitem[{Cao et~al.(2009)Cao, Jia and Liu}]{CaoEtal09}
\bibinfo{author}{Cao\xfnm[ L.]}, \bibinfo{author}{Jia\xfnm[ Z.]},
  \bibinfo{author}{Liu\xfnm[ J.]}.
\newblock \bibinfo{title}{Computation of medial axis and offset curves of
  curved boundaries in planar domains based on the cesaro¡¯s approach}.
\newblock \bibinfo{journal}{Computer Aided Geometric Design}
  \bibinfo{year}{2009};\bibinfo{volume}{26}:\bibinfo{pages}{444--454}.
\bibitem[{Barnhill et~al.(1987)Barnhill, Farin, Jordan and
  Piper}]{BarnhillEtal87}
\bibinfo{author}{Barnhill\xfnm[ R.E.]}, \bibinfo{author}{Farin\xfnm[ G.]},
  \bibinfo{author}{Jordan\xfnm[ M.]}, \bibinfo{author}{Piper\xfnm[ B.R.]}.
\newblock \bibinfo{title}{Surface/surface intersection}.
\newblock \bibinfo{journal}{Computer Aided Geometric Design}
  \bibinfo{year}{1987};\bibinfo{volume}{4}:\bibinfo{pages}{3--16}.

\end{thebibliography}

\end{document}